\documentstyle[12pt]{article}
\newlength{\dinwidth}
\newlength{\dinmargin}
\begin{document}
\begin{center}
{\Large \bf
DETERMINATION OF THE  ELECTROWEAK CHIRAL-LAGRANGIAN PARAMETERS AT THE
LHC} \vskip 1cm
{\large \bf A. Dobado}  \footnote{E-mail: dobado@cernvm.cern.ch}    \\
{\large Departamento de F\'{\i}sica Te\'orica  \\
 Universidad Complutense de Madrid\\
 28040 Madrid, Spain \\
 and  \\}
{\large \bf M. T. Urdiales} \footnote {E-mail: mayte@vm1.sdi.uam.es } \\
{\large Departamento de F\'{\i}sica Te\'orica  \\
 Universidad Aut\'onoma de Madrid\\
 28049 Madrid, Spain \\}
\vskip 1cm
February 1995 \\
\vskip 1cm
\begin{abstract}
In this work we report on the results obtained in a detailed
and systematical study of the possibility to measure the parameters
appearing in the electroweak chiral lagrangian. The
main novelty of our approach  is that we do not use the Equivalence Theorem
and therefore we work explicitly with all the gauge boson degrees of freedom.
\end{abstract}
\end{center}
\vskip 1.0cm
hep-ph/9502255\\
\newpage
\textheight 20 true cm

\section{Introduction}

Today it is clear for many physicists that one of the main goals of the
future CERN Large Hadron Collider ($LHC$) is to find as much information
as
possible about the nature of the Standard Model ($SM$) Electroweak Symmetry
 Breaking ($ESB$)
 \cite{M.E.}. In spite of the huge amount of data
obtained in  the last years
at the Large Electron-Positron Collider ($LEP$) it is very few what we really
know about the $ESB$. The proposed mechanisms include   ideas such as
supersymmetry (see \cite{Susy} and references therein), technicolor
\cite{Techni} and many others. Therefore it would
 be very
interesting to have some model independent framework to make a
phenomenological description of the $ESB$ physics.

In fact such a
framework exists, at least for the strongly interacting case i.e., when no
light modes are present in the $ESB$. It has been developed in the last
years and used to describe the scattering of the longitudinal
components ($LC$) of the electroweak gauge bosons \cite{HeE1} as well as the
precision tests of the Standard Model coming from $LEP$ \cite{HeE2}. It
is based
on the application of the chiral Lagrangians or Chiral Perturbation
Theory
($\chi PT$ )\cite{Wleff} (previously invented for
 the description of the low-energy hadron interactions) to the dynamics of the
$ESB$ Goldstone bosons ($GB$).

One assumes that there is a physical
system with a global symmetry group $G$ which is spontaneously broken to
some other group $H$. This global symmetry breaking drives the gauge
electroweak
$SU(2)_L \times U(1)_Y$ symmetry breaking down
to $U(1)_{em}$
through the well known Higgs mechanism. The only election for the $G$
and $H$ groups compatible with the presence of the $SU(2)_{L+R}$
custodial symmetry \cite{cus} (to have a $\rho$ parameter naturally close to
one) and the existence of three massive gauge bosons after the $ESB$
i.e., the $W^+$, $W^-$ and $Z^0$, is $G=SU(2)_L \times SU(2)_R$,
$H=SU(2)_{L+R}$.

Then the low energy dynamics of the $ESB$ is described
by a $SU(2)_L \times U(1)_Y$ gauged non-linear sigma model ($GNLSM$)
including an arbitrary large number of terms in the action with
different number of derivatives of the $GB$ fields and electroweak gauge
bosons. The corresponding couplings (parameters) encode the dynamics of
the $ESB$ sector of the $SM$ and must be renormalized to absorb
divergences. However, at low enough energies only a small number of
terms (and couplings) are needed to eliminate all the divergences.
In principle the
values of these couplings or parameters could be obtained from the
underlying theory or directly by fitting them from future experiments.

In this work we will study the possibilities of the $LHC$ for measuring
these parameters and the expected corresponding errors. In particular we
will concentrate in events producing $Z^0Z^0$ or $W^{\pm}Z^0$ pairs.
As it was mentioned above, the application
of the chiral Lagrangian technique to the production of electroweak
gauge bosons is not new and has already been considered at the
literature \cite{si}. However, all the applications worked out until now are
based in the so called Equivalence Theorem ($ET$) \cite {ET}.
This theorem relates the
$S$ matrix elements of processes containing  electroweak gauge
bosons $LC$ with the corresponding processes with $GB$. However, in
a recent work \cite{ETCL} concerning the formulation of the $ET$ in the
context of $\chi PT$, the Equivalence Theorem is
severely restricted to a narrow energy applicability window. It
can also be applied in the high energy domain together with the $\chi PT$ but
using some non-perturbative technique, like dispersion relations \cite{Uni}
or the large $N$ limit \cite{largeN}, in order to have an appropriate
unitarity behaviour of the amplitudes.

For this reason we consider prioritary to apply directly the chiral
Lagrangian description of the $ESB$ without using the $ET$. The main
problems of this approach are two: First one has to include
explicitly
the gauge degrees of freedom in the model which makes the computations
extremely more difficult. Second, one has to restrict the results to the
low energy region where standard $\chi PT$ can safely be applied thus
losing many higher energy events. The advantage is that the values of
the fitted parameters will be more reliable since one is not using
the $ET$ complemented with some non-perturbative method.

The plan of this work goes as follows: In section 2 we introduce the
chiral effective lagrangian to be used with its parameters and the
corresponding Feynman rules. In section 3 we compute the cross-section
of the $LHC$ subprocesses that are relevant for measuring the
parameters. In section 4 we consider the signatures and the possible
backgrounds. In section 5 we show how we compute the total number of
expected events at the $LHC$ from the subprocess cross-sections for the
signals and backgrounds. In section 6 we discuss the sensitivity of the
machine to the different parameters, we define the optimal cuts and we compute
the statistical
significance of the different parameter measures and estimate the statistical
errors. In
section 7 we consider other sources of systematical errors such as the
uncertainty
on
the proton structure functions and in section 8 the effect of the running of
the
parameters.
Finally, in section 9 we review the main conclussions of our work.

\section{The effective Lagrangian and the Feynman rules}

In this section we consider the most general effective Lagrangian for
the $ESB$
compatible with the $SU(2)_L \times U(1)_Y$ local symmetry of the
$SM$ and the breaking pattern $SU(2)_L \times U(1)_Y
\rightarrow U(1)_{\rm em}$. The well known chiral Lagrangian describing
the $ESB$ of the $SM$, can be written as an infinite
expansion with
terms of increasing number of gauge fields and derivatives of the Goldstone
bosons ($GB$), with an infinite number of arbitrary parameters. This chiral
Lagrangian can be seen as a low momentum expansion for the corresponding
Green functions. At  some given order in the
number of  $GB$ derivatives
one can work only with a finite number of terms and  parameters. In
this case the model can only be applied to  much smaller energies than
 $4\pi v$
 which is the
 parameter controlling this expansion ($4\pi v
\simeq 3 TeV$ since $v\simeq 250 GeV$).

The first term of this effective Lagrangian ($O(p^2)$) is given by:

\begin{equation}
{\cal L}^{(2)}=\frac{v^2}{4}tr[D_{\mu} U(D^{\mu}U)^{\dagger}]
\label{eq:lag1}
\end{equation}

and by the Yang-Mills Lagrangian:

\begin{equation}
{\cal L}_{YM}=-\frac{1}{4} B_{\mu\nu} B^{\mu\nu} - \frac{1}{2}
Tr[F_{\mu\nu}F^{\mu\nu}]
\label{eq:YM}
\end{equation}

We choose the GB parametrization as an unitary matrix $U$ belonging
to the quotient space $SU(2)_L \times SU(2)_R / SU(2)_{L+R}$.

\begin{displaymath}
U = {\rm exp} (i \frac{\vec{\tau}.\vec{\pi}}{v})
\end{displaymath}

where
$\vec{\tau} \equiv ({\tau}_1, {\tau}_2,{\tau}_3)$ are the Pauli
matrices and $\vec{\pi} \equiv ({\pi}_1(x), {\pi}_2(x), {\pi}_3(x))$
represents the triplet of $GB$.

The covariant derivative $D_{\mu}U$ is defined as:

\begin{displaymath}
D_{\mu}U = {\partial}_{\mu} U + ig W_{\mu} U - ig' U Y_{\mu}
\end{displaymath}

Here, $W_{\mu}$ and $Y_{\mu}$ are the $SU(2)_L$ and the $U(1)_Y$ gauge
fields given by:

\begin{displaymath}
W_{\mu} = \frac{{\vec{W}}_{\mu} {\vec{\tau}}}{2}
\end{displaymath}

\begin{displaymath}
Y_{\mu} = \frac{B_{\mu} {\tau}_3}{2}
\end{displaymath}

where $W_{\mu}=(W^1_{\mu},W^2_{\mu},W^3_{\mu})$ represents the triplet
of $SU(2)_L$ gauge fields.

As usual, the covariant field strength
tensors are defined as

\begin{eqnarray*}
F_{\mu\nu}(x) & = & {\partial}_{\mu} W_{\nu}(x) - {\partial}_{\nu}
W_{\mu}(x) + ig [W_{\mu}(x),W_{\nu}(x)]  \\
B_{\mu\nu}(x) & = & {\partial}_{\mu} B_{\nu}(x) - {\partial}_{\nu}
B_{\mu}(x)
\end{eqnarray*}

The transformation properties under the $SU(2)_L \times U(1)_Y$ gauge
transformations are

\begin{eqnarray*}
ig W'_{\mu} & = & g_L (x) ig W_{\mu} g_L^{\dagger} (x) +  g_L (x)
{\partial}_{\mu} g_L^{\dagger} (x)  \\
ig' Y'_{\mu} & = & ig' Y_{\mu} + g_Y (x) {\partial}_{\mu}
g_Y^{\dagger} (x) \\
U' & = & g_L (x) U {g_Y}^{\dagger} (x)
\end{eqnarray*}

where

\begin{eqnarray*}
g_L(x) & = & e^{i {\theta}_k (x) {\tau}^{k} /2}    \\
g_Y(x) & = & e^{i \beta(x).{\tau}^3 /2}
\end{eqnarray*}

Note that, as expected, the
$GB$ fields, ${\pi}_i$, transform non-linearly .

In order to construct the chiral lagrangian to order  $O(p^4)$ we define,
following Longhitano in \cite{Long}, the quantities $T$, $V_{\mu}$ and
 ${\cal D}_{\mu} O(x)$:

\begin{eqnarray*}
T & = & U {\tau}^3 U^{\dagger}  \\
V_{\mu} & = & D_{\mu}U U^{\dagger}   \\
{\cal D}_{\mu} O(x) & = & {\partial}_{\mu} O(x) + ig [W_{\mu}(x), O(x)]
\end{eqnarray*}

Then the complete electroweak chiral Lagrangian with the whole set of $SU(2)
\times U(1)_Y$, Lorentz, $C$, $P$ and $T$ invariant operators up to
dimension four has the form:

\begin{equation}
{\cal L}={\cal L}^{(2)}+{\cal L}_{YM}+{\cal L'}_1+ \sum_{i=1}^{13} {\cal
L}_i
\label{eq:lag3}
\end{equation}

In this equation, the different terms ${\cal L}_i$ are the $SU(2)_L
\times U(1)_Y$ invariant functions of the gauge vector bosons and the $GB$
fields, containing  four derivatives, whereas ${\cal L}'_1$ has
dimension two. They have the following expressions:

\begin{eqnarray}
{\cal L'}_1  & = & \frac{1}{4} g^2 {\alpha}_0 v^2 [Tr(TV_{\mu})]^2
\nonumber   \\
{\cal L}_1  & = & \frac{1}{2} g^2 {\alpha}_1 B_{\mu\nu} Tr(TF^{\mu\nu})
\nonumber   \\
{\cal L}_2  & = & \frac{1}{2} i g {\alpha}_2 B_{\mu\nu}
Tr(T[V^{\mu},V^{\nu}])  \nonumber   \\
{\cal L}_3  & = & i g {\alpha}_3 Tr(F_{\mu\nu}[V^{\mu},V^{\nu}])
\nonumber   \\
{\cal L}_4  & = & {\alpha}_4 [Tr(V_{\mu}V_{\nu})]^2   \nonumber  \\
{\cal L}_5  & = & {\alpha}_5 [Tr(V_{\mu}V^{\mu})]^2   \nonumber  \\
{\cal L}_6  & = & {\alpha}_6 Tr[(V_{\mu}V_{\nu})]
Tr(TV^{\mu}) Tr(TV^{\nu})  \nonumber    \\
{\cal L}_7  & = & {\alpha}_7 Tr[(V_{\mu}V^{\mu})] [Tr(TV^{\nu})]^2
\nonumber   \\
{\cal L}_8  & = & \frac{1}{4} g^2 {\alpha}_8 [Tr(TF_{\mu\nu})]^2
\nonumber   \\
{\cal L}_9  & = & \frac{1}{2} i g {\alpha}_9 Tr(TF_{\mu\nu})
Tr(T[V^{\mu},V^{\nu}])  \nonumber   \\
{\cal L}_{10}  & = & \frac{1}{2} {\alpha}_{10}
[Tr(TV_{\mu})Tr(TV_{\nu})]^2     \nonumber  \\
{\cal L}_{11} & = & {\alpha}_{11} Tr[({\cal D}_{\mu} V^{\mu})^2]
\nonumber  \\
{\cal L}_{12} & = & \frac{1}{2} {\alpha}_{12} Tr(T{\cal D}_{\mu}{\cal
D}_{\nu}V^{\nu}) Tr(TV^{\mu})  \nonumber  \\
{\cal L}_{13} & = & \frac{1}{2} {\alpha}_{13} [Tr(T{\cal D}_{\mu}
V^{\nu})]^2
\label{eq:lag2}
\end{eqnarray}

As it is well known there is some arbitrariness in the choice of a
particular base of invariants. Using the equations of motion it would be
possible to eliminate the operators ${\cal L}_{11}$, ${\cal L}_{12}$
and ${\cal L}_{13}$  redefining the rest of the terms (see for
example \cite{Fer}).

So, a set of $\alpha_k$ ($k=0,..13$) parameters have
appeared in the definition of this effective Lagrangian.
We display in Table \ref{tab:equiv} the relations between some different
sets of the chiral Lagrangian parameters used in the literature, i.e. by
Longhitano \cite{Long}, Feruglio \cite{Fer} and Dobado et.al.
\cite{HeE2}.

The theory defined by the Lagrangian above is non-renormalizable in the
strict sense, but the divergences appearing when we calculate at
one-loop with the chiral Lagrangian at lowest order eq.(\ref{eq:lag1})
have the same form that some of the terms  obtained  at tree level in
the following order in the chiral expansion. The counterterms needed to
reabsorb the divergences generated to the one-loop level with ${\cal
L}^{(2)}$ in the Landau gauge were obtained by the authors in
\cite{Long}.
They have the same form that ${\cal L}'_1$, ${\cal L}_1$, ${\cal L}_2$,
${\cal L}_3$, ${\cal L}_4$ and ${\cal L}_5$, whereas  the rest of
the terms in eq.(\ref{eq:lag2}) are not needed for renormalization.

We will describe the interactions between gauge bosons and fermions
by the same Lagrangian than that of the $SM$, which can also describe
the couplings between fermions and scalars by means of the usual
Yukawa terms.
However the latter are not used in our calculations since we work all
the time with massless fermions.

 Once we have set the Lagrangian to be used, the next step is to define
the quantum
theory. This can be done in a standard way  using the Faddeev-Popov
method and choosing some appropriate  $R_{\xi}$ covariant gauge. The gauge
fixing
functions are constructed by requiring  that the terms of the form $W^{i}_{\mu}
{\partial}^{\mu} {\pi}_i(x)$ and $B_{\mu} {\partial}^{\mu} {\pi}_3$ cancel
out; thus yielding:

\begin{eqnarray}
f_i(W_{i\mu})  & =  & {\partial}^{\mu} W_{i\mu} - \frac{v}{2} {\xi}_W
g{\pi}_i,
\;\;\;\;\;  i=1,2,3     \nonumber  \\
f_0(B_{\mu})  & =  & {\partial}^{\mu} B_{\mu}  + \frac{v}{2} {\xi}_B g'
{\pi}_3
\end{eqnarray}

Now we can write the gauge-fixing and Faddeev-Popov terms as:

\begin{displaymath}
{\cal L}_{GF} = - \frac{1}{2 {\xi}_W} \sum_{i=1}^{3} [f_i(W_{i\mu})]^2 -
\frac{1}{2 {\xi}_B} [f_0(B_{\mu})]^2
\end{displaymath}

\begin{eqnarray*}
{\cal L}_{FPG} & = & \int d^4 y {c^{\dag}}_m(x) \frac{\delta
f_m(x)}{\delta {\theta}_n (y)} c_n(y)  \;\;\;\;\;\;  m,n= 0,1,2,3
\end{eqnarray*}

Therefore, the effective action and the effective Lagrangian
(non-linear $SM$ Lagrangian, ${\cal L}_{NLSM}$) take the form:

\begin{equation}
S_{eff} =  \int d^4x [ {\cal L}(x) + {\cal L}_{GF} + {\cal L}_{FPG}]
\end{equation}

\begin{equation}
{\cal L}_{NLSM} = {\cal L}^{(2)}+{\cal L'}_1+{\cal
L}^{(4)}+{\cal L}_{YM}+{\cal L}_{GF}
+{\cal L}_{FPG}+{\cal L}_{\Psi}
\label{eq:MENL}
\end{equation}

In the following  we will to work in the Landau gauge
(${\xi}_W={\xi}_B=0$). In this gauge
the perturbative $\pi$ propagator remains massless and the
Faddeev-Popov ghosts only couple to the gauge field. There is not direct
coupling of the $\pi$ field to the ghosts. Moreover, the counterterms necessary
to cancel divergences with ${\cal L}^{(2)}$ at one loop are gauge
invariant functions of the gauge fields and the $GB$ alone (the ${\cal L}'_1$,
 ${\cal L}_1$, ${\cal L}_2$,
${\cal L}_3$, ${\cal L}_4$ and ${\cal L}_5$ terms mentioned
above). In other gauges, or other parametrizations of the coset
space, the counterterms could also be functions of the ghost fields, and their
 structure would be determined using the Becchi-Rouet-Stora ($BRS$)
invariance.

Thus, the gauge fixing and the Faddeev-Popov term in the Landau gauge read:

\begin{eqnarray}
{\cal L}_{GF} & = & - \frac{1}{2 {\xi}} \sum_{i=1}^{3}
[{\partial}^{\mu}(W_{i\mu})]^2 - \frac{1}{2 {\xi}}
[{\partial}^{\mu}(B_{\mu})]^2  \nonumber   \\
{\cal L}_{FPG} & = & {\partial}_{\mu} {c_i}^{\dag} {\partial}^{\mu} c^i
- g{\epsilon}_{ijk} {c^i}^{\dag} c^j {\partial}^{\mu} {W_{\mu}}^k +
{\partial}_{\mu} c_0^{\dag} {\partial}^{\mu} c^0
\label{eq:quant}
\end{eqnarray}

In the Appendix we show the Feynman rules
derived from the Lagrangian ${\cal L}_{NLSM}$ eq.(\ref{eq:MENL}) that we
use for our calculations.
They correspond to Figures \ref{fig:GF2}, \ref{fig:GF3} and
\ref{fig:GF4}. As usual it has
been useful to redefine the gauge fields as:

\begin{eqnarray*}
W^{\pm}_{\mu} & = & \frac{{W_{\mu}}^1 \mp i{W_{\mu}}^2}{\sqrt{2}}  \\
Z_{\mu} & = & {\rm cos}{\theta}_W {W_{\mu}}^3 -{\rm sin}{\theta}_W
B_{\mu}
\\ A_{\mu} & = & {\rm sin}{\theta}_W {W_{\mu}}^3 + {\rm cos}{\theta}_W
B_{\mu}
\end{eqnarray*}

where the weak angle is defined by

\begin{displaymath}
{\rm tg}{\theta}_W = \frac{g'}{g}
\end{displaymath}

and  the $GB$ as:

\begin{eqnarray*}
{\pi}^{\pm} & = & \frac{{\pi}^1 \mp i{\pi}^2}{\sqrt{2}}  \\
{\pi}^0 & = & {\pi}^3
\end{eqnarray*}

\section{The subprocesses cross-sections}

In this work we are interested in studying of the different cross
sections that
contribute to the final states $Z^0Z^0$ and $W^{\pm}Z^0$ at $pp$
colliders such as $LHC$ since they are the most promising from the experimental
point of view. In both cases we focus our attention on the so-called
gold-plated
events, where the produced bosons decay to the leptonic final states
$l=e,\mu,
\nu$. In spite of the very small branching ratios ($BR$),

\begin{eqnarray*}
BR(Z^0Z^0 \rightarrow l \bar{l} l'\bar{l'}) & = & 0.0044   \\
BR(W^{\pm}Z^0 \rightarrow l {\nu}_{l} l'\bar{l'}) & = & 0.013
\end{eqnarray*}

these are the most
interesting events since they are much easier to detect than the hadronic
channels.

On the other hand, we will assume that it is not possible to
detect experimentally the different
polarizations of the gauge bosons in
the final state. Therefore we have to include in our computations all the
$Z^0$ or $W^{\pm}$ gauge bosons, either transversally or longitudinally
polarized. As it was mentioned in the introduction, instead of
using the $ET$  to calculate the
collisions of the weak bosons, as it is usually done, we have computed
the amplitudes with gauge bosons as initial and final states and then
projected them
in all their polarizations. Thus, as we are not going to apply the
$ET$ we only consider the maximal bound on the energy required for the
applicability of the chiral Lagrangian formalism, but we do not have any low
energy bound. In this way we can take into account the influence of the chiral
lagrangian parameters in all the gauge bosons polarization channels
and not only in the longitudinal ones as it is usually the case. The main
disadvantage in our procedure is that the calculations become much more
complicated. This fact leads us to work only at tree level to make the
computation more accessible.

In principle, according to the spirit of $\chi PT$,
the one-loop corrections coming from the lowest order lagrangian should also be
included since they are $O(p^4)$. However, the computation of these corrections
is extremely involved. In addition, they have not dependence on the chiral
parameters. As it will become clear later, our main
interest is to study the dependence of the number of
events on the chiral parameters in order to see which of them could
be measured
at the $LHC$. For this reason we do not expect that our results
concerning
the measurable parameters would change too much when the
one-loop
corrections are included but, of course, the precise number of events
will do.

Later we will also study the effect of the running of the parameters on
our results
which in some way takes into account part of the one-loop contribution, but, in
any case, our approach is the simplest one including the right
dependence on the chiral lagrangian parameters.

Now we describe how we have calculated the cross sections of the above
mentioned processes. From the quantized chiral Lagrangian that appears
in eq.(\ref{eq:MENL}) we obtain the corresponding
Feynman rules (see Appendix). Then we compute, at tree level, the
amplitudes for the different
subprocesses contributing to the $Z^0Z^0$ or $W^{\pm}Z^0$ final state,
using a REDUCE code, and then the corresponding cross-sections.
We calculate them in the center of mass frame, with a total energy
$\sqrt{\hat{s}}$. There, the four
momenta of the interacting particles are:

\begin{eqnarray*}
p_1 & = & (E_1, 0,0,p) \\
p_2 & = & (E_2, 0,0,p) \\
p_3 & = & (E_3,  p'{\rm sin}\theta , 0,  p'{\rm cos}\theta)   \\
p_4 & = & (E_4, -p'{\rm sin}\theta , 0, -p'{\rm cos}\theta)
\end{eqnarray*}

Here, $E_i$ are the particle energies, and $p$ and  $p'$ are,
respectively,
the magnitudes of the three-momenta of the initial and the final states.

\begin{eqnarray*}
p & = & \frac{1}{2\sqrt{\hat{s}}}{[{\hat{s}}^2
-2{\hat{s}}({m_1}^2+{m_2}^2)+({m_2}^2-{m_1}^2)^2]}^{1/2}  \\
p' & = & \frac{1}{2\sqrt{\hat{s}}}{[{\hat{s}}^2
-2{\hat{s}}({m_3}^2+{m_4}^2)+({m_4}^2-{m_3}^2)^2]}^{1/2}  \\
E_i & = & \sqrt{{m_i}^2+{p_i}^2} \;\;\;\;\;   i=1,..4  \\
{\hat{s}} & = & (p_1+p_2)^2    \\
{\hat{t}} & = & (p_1-p_3)^2    \\
{\hat{u}} & = & (p_1-p_4)^2
\end{eqnarray*}

where $\hat{s}$, $\hat{t}$ and $\hat{u}$ are the Mandelstam variables.
The differential cross section is given by this expression:

\begin{equation}
d {\hat{\sigma}} = \frac{1}{32 \pi {\hat{s}}} \frac{p'}{p} \sum {\mid M
\mid}^2 d{\rm cos} {\theta}
\label{eq:subpro}
\end{equation}

where $M$ is the helicity amplitude, and the $\Sigma$ symbol refers to the sum
of all the final polarizations and the average on the
initial ones. In the case of electroweak gauge bosons scattering, where
we separate the transversal and longitudinal
polarizations in the initial state (as we will see in section 5), we
take into account
the different helicities contributing to $TT$, $TL$, $LT$ and $LL$
polarizations of the initial bosons, to do this average.
We also
include the corresponding branching ratios to the gold-plated events.

Let us consider now the processes corresponding to the elastic collision of
gauge bosons (fusion processes). Symbolically this reaction can be written as:

\begin{displaymath}
V_1 (p_1,{\lambda}_1) V_2 (p_2,{\lambda}_2) \rightarrow
V_3 (p_3,{\lambda}_3) V_4 (p_4,{\lambda}_4)
\end{displaymath}

We choose the following polarization vectors in the helicity base
(${\lambda}_i=\pm 1,0$):

\begin{eqnarray*}
{\epsilon}^{\pm}_1 & = & \frac{1}{\sqrt{2}} (0,{\mp}1, -{\rm i},0)   \\
{\epsilon}^0_1  & = & \frac{1}{m_{V_1}} (p,0,0,E_1)     \\
{\epsilon}^{\pm}_2 & = & \frac{1}{\sqrt{2}} (0,{\mp}1,  {\rm i},0)   \\
{\epsilon}^0_2  & = & \frac{1}{m_{V_2}} (p,0,0,-E_2)     \\
{\epsilon}^{\pm}_3 & = & \frac{1}{\sqrt{2}} (0, \mp {\rm cos} \theta,
 -{\rm i}, \pm {\rm sin}\theta) \\
{\epsilon}^0_3 & = & \frac{1}{m_{V_3}} (p', E_3 {\rm sin} \theta,
0,  E_3 {\rm cos} \theta)  \\
{\epsilon}^{\pm}_4 & = & \frac{1}{\sqrt{2}} (0, \mp {\rm cos} \theta,
 {\rm i}, \pm {\rm sin} \theta)  \\
{\epsilon}^0_4 & = & \frac{1}{m_{V_4}} (p', -E_4 {\rm sin} \theta,
0, -E_4 {\rm cos} \theta)
\end{eqnarray*}

The number of independent helicity amplitudes in fusion processes of
massive gauge bosons, such as $Z^0Z^0 \rightarrow
Z^0Z^0$, $W^+W^- \rightarrow Z^0Z^0$, $W^+Z^0 \rightarrow
W^+Z^0$ or $W^-Z^0 \rightarrow W^-Z^0$,
is $3^4=81$. The photon has only two polarized
transverse states and therefore, in the processes $W^+\gamma
\rightarrow W^+Z^0$ or $W^-\gamma \rightarrow W^-Z^0$
the number of helicity amplitudes is $2 \times 3^3 = 54$.
As our chiral Lagrangian  affecting the boson scattering is
invariant under
$C$, $P$ and $T$ transformations, we can derive many relations between
different
helicity amplitudes. However we have not
used those relations to save computations. Instead we have calculated all the
helicity amplitudes in each process and the relations
between them have been only used to check our results.

\subsection{$Z^0Z^0$ final state}

Now we concentrate in the different processes  that
contribute to the final state $Z^0Z^0$ \cite{si}:

\begin{eqnarray*}
gg           &   \rightarrow   &   Z^0Z^0  \\
q\bar{q}     &   \rightarrow   &   Z^0Z^0  \\
Z^0_TZ^0_T   &   \rightarrow   &   Z^0Z^0  \\
Z^0_TZ^0_L   &   \rightarrow   &   Z^0Z^0  \\
Z^0_LZ^0_T   &   \rightarrow   &   Z^0Z^0  \\
Z^0_LZ^0_L   &   \rightarrow   &   Z^0Z^0  \\
W^+_TW^-_T   &   \rightarrow   &   Z^0Z^0  \\
W^+_TW^-_L   &   \rightarrow   &   Z^0Z^0  \\
W^+_LW^-_T   &   \rightarrow   &   Z^0Z^0  \\
W^+_LW^-_L   &   \rightarrow   &   Z^0Z^0
\end{eqnarray*}

Note that we separate  the contribution of the different polarizations
channels in the initial state, since different
polarizations will
have different luminosities in $pp$ collisions. In
particular we use the
Weizsaker-Williams  \cite{fegamm}  and the effective $W$ approximation
\cite{Daw} to compute the
$V_1V_2$ pair luminosity  in the $LHC$ beams. Therefore, we have to
divide
 these cross-sections
into the contributions coming from the different gauge
boson initial polarizations.

On the other hand, it is well known that at supercollider energies the
one-loop
process $gg \rightarrow ZZ$ is not negligible. The gluon-gluon fusion cross
section was calculated  in the Minimal Standard
Model, $MSM$ (with just one Higgs doublet) via one-loop of quarks
by Glover and Van deer Bij \cite{Glo}. The
corresponding diagrams are shown in Figure \ref{fig:gg}. As we are
using a chiral Lagrangian
description of the ESB sector we do not include the contribution coming from
diagrams with the Higgs boson in our computations.

The quark-antiquark annihilation represents the main source of
$Z^0Z^0$ pairs in $pp$ colliders like the $LHC$. As in the case of
gluon fusion,
there is no dependence on the chiral parameters in the calculation, at our
level of approximation. The cross
section at tree level, which only receives contribution  from the
$t$ and $u$ channels (Figure \ref{fig:qq}), has the same well-known
expression as in the $MSM$.

The $Z^0Z^0$ fusion calculated using the chiral Lagrangian,
only receives contribution from the vertex $Z^0Z^0Z^0Z^0$
(Figures \ref{fig:GF4} and \ref{fig:ZZ}).
The only dependence on the effective Lagrangian parameters is through
the following combination eq.(\ref{eq:GF4}):

\begin{displaymath}
{\alpha}_4 + {\alpha}_5 + 2({\alpha}_6 + {\alpha}_7 + {\alpha}_{10})
\end{displaymath}

Again we can use the relations derived from the $C$, $P$ and
$T$ to relate the helicity
amplitudes and thus to check our results. These relations are:

\begin{eqnarray*}
|M_{{{\lambda}_1}{{\lambda}_2}{{\lambda}_3}{{\lambda}_4}}| &
\stackrel{C}{=}  &
|M_{{{\lambda}_2}{{\lambda}_1}{{\lambda}_4}{{\lambda}_3}}|        \\
|M_{{{\lambda}_1}{{\lambda}_2}{{\lambda}_3}{{\lambda}_4}}| &
\stackrel{P}{=}  &
|M_{{-{\lambda}_1}{-{\lambda}_2}{-{\lambda}_3}{-{\lambda}_4}}|    \\
|M_{{{\lambda}_1}{{\lambda}_2}{{\lambda}_3}{{\lambda}_4}}| &
\stackrel{T}{=}  &
|M_{{{\lambda}_3}{{\lambda}_4}{{\lambda}_1}{{\lambda}_2}}|
\end{eqnarray*}
In order to relate some helicity amplitudes
we have also taken into account whether the scattering particles are
identical or not.,
There are still $15$ independent amplitudes remaining after applying these
symmetry relations.

The $W^+W^- \rightarrow Z^0Z^0$ reaction represents another
source of $Z^0Z^0$ pairs from $pp$ beams. Every
helicity amplitude in this process receives contributions from the
$t$ and $u$ channels, by exchanging a  $W$ or a
$\pi$, and directly from the vertex $W^+W^-Z^0Z^0$
(Figure \ref{fig:WW}). Therefore we can write:

\begin{displaymath}
M_{{{\lambda}_1}{{\lambda}_2}{{\lambda}_3}{{\lambda}_4}} = [M_{t_1} +
M_{t_2} + M_{u_1} + M_{u_2} + M_4 ]_{{{\lambda}_1}{{\lambda}_2}
{{\lambda}_3}{{\lambda}_4}}
\end{displaymath}

where the subscript $1$($2$) is referred to the exchange of a gauge
boson
(Goldstone boson) through the corresponding channel. There are $81$
helicity amplitudes, but the number of independent $M_{{{\lambda}_1}
{{\lambda}_2} {{\lambda}_3} {{\lambda}_4}}$ is reduced to $25$ by means of
symmetry relations derived from   $P$ and $C$ invariance.

As we expected, the amplitudes that we have obtained satisfy these
equalities.
On the other hand,
all the chiral parameters, ${\alpha}_k$ (but  ${\alpha}_{10}$),
affect this process as we can deduce from the Feynman rules
eqs.(\ref{eq:GF3} and \ref{eq:GF4})
and from the different diagrams that contribute to the collision
$W^+W^- \rightarrow Z^0Z^0$.

\subsection{$W^{\pm}Z^0$ final state}

As it was previously said, we have also studied the $W^{\pm}Z^0$ final
state. The different sources for  $W^{\pm}Z^0$ pairs in $pp$
colliders at tree level order are the following:

\begin{eqnarray*}
q\bar{q'}  &  \rightarrow  &  W^+Z^0    \\
q'\bar{q}  &  \rightarrow  &  W^-Z^0    \\
W^+_T Z^0_T &  \rightarrow  &  W^+ Z^0  \\
W^+_T Z^0_L &  \rightarrow  &  W^+ Z^0  \\
W^+_L Z^0_T &  \rightarrow  &  W^+ Z^0  \\
W^+_L Z^0_L &  \rightarrow  &  W^+ Z^0  \\
W^-_T Z^0_T &  \rightarrow  &  W^- Z^0  \\
W^-_T Z^0_L &  \rightarrow  &  W^- Z^0  \\
W^-_L Z^0_T &  \rightarrow  &  W^- Z^0  \\
W^-_L Z^0_L &  \rightarrow  &  W^- Z^0  \\
W^+_T \gamma & \rightarrow  &  W^+ Z^0  \\
W^+_L \gamma & \rightarrow  &  W^+ Z^0  \\
W^-_T \gamma & \rightarrow  &  W^- Z^0  \\
W^-_L \gamma & \rightarrow  &  W^- Z^0
\end{eqnarray*}

where we call $q=u,c$ and $q'=d,s,b$.

Most of the boson pairs $W^{\pm}Z^0$ are obtained in $pp$ colliders
via quark-antiquark annihilations. Using the chiral Lagrangian framework
we calculate  the corresponding cross section at tree level order.
As we
can see in Figure \ref{fig:qq'}, three standard $s$, $t$ and $u$
diagrams contribute,
but the new physics coming from the non-linear lagrangian is isolated in
the three boson vertex, which in this process affects only the $s$
channel.

\begin{displaymath}
M = M_s + M_t + M_u
\end{displaymath}

If we look at the Feynman rules (Figure \ref{fig:GF3}) we can deduce
that the set of
${\alpha}_k$ parameters contributing to this process are ${\alpha}_1$,
${\alpha}_2$, ${\alpha}_3$, ${\alpha}_8$, ${\alpha}_9$, ${\alpha}_{11}$,
${\alpha}_{12}$ and ${\alpha}_{13}$ eq.(\ref{eq:GF3}). In the hadronic
collider case, the three boson vertex has been studied in this process
with chiral Lagrangians, in \cite{Luk} at tree level,  using the $ET$
(by J.Bagger, et.al. in  \cite{si}), or in our previous study
\cite{AnMa}
for the case $g'=0$ and including the running of the
couplings. In this last case we analyzed the sensitivity to
the ${\alpha}_3$ parameter. Here we work firstly at tree level order, but
including the dependence on the $\alpha_k$ parameters in all the
channels to produce $W^{\pm}Z^0$ pairs in $pp$ colliders. Finally we
will include the running parameters effect in section 8.

Other mechanism to obtain $W^{\pm}Z^0$ pairs is through
  $W^{\pm}Z^0$ or $W^{\pm} \gamma$  collisions transversally or
longitudinally polarized at the initial state. In both cases the different
polarization amplitudes are obtained by adding the contribution of the $s$
and $u$ channels and the four gauge boson vertex in
(Figures \ref{fig:WZ} and \ref{fig:Wf}).
The exchanged
particles  in the $s$ and $u$ diagrams are a gauge boson ($s_1$,
$u_1$) and a $GB$ ($s_2$, $u_2$), respectively.

\begin{displaymath}
M_{{{\lambda}_1}{{\lambda}_2}{{\lambda}_3}{{\lambda}_4}} = [M_{s_1} +
M_{s_2} + M_{u_1} + M_{u_2} + M_4 ]_{{{\lambda}_1}{{\lambda}_2}
{{\lambda}_3}{{\lambda}_4}}
\end{displaymath}

In the $W^{\pm}Z^0 \rightarrow  W^{\pm}Z^0$ collisions  there are
initially $2 \times 3^4$ helicity amplitudes. This $2$ factor can be
dropped  using $C$ invariance that relates the magnitude of the
helicity amplitudes
$M_{{\lambda}_1{\lambda}_2{\lambda}_3{\lambda}_4}$ in the
process with $W^+$ and with $W^-$. If we apply the relations between
amplitudes derived from $P$ and $T$ invariance, there will be only
$25$ remaining amplitudes.
These subprocesses cross sections are affected by all the
${\alpha}_k$, but  ${\alpha}_{10}$, as we can deduce
from the Feynman rules eqs.(\ref{eq:GF3} and \ref{eq:GF4}).

The number of helicity amplitudes corresponding to
the processes $W^{\pm}\gamma  \rightarrow  W^{\pm}Z^0$ is $2 \times (2
\times 3^3)$. The $2$ factor disappears when we take into account $C$
invariance as in the previous case. The number of remaining amplitudes is
reduced again, by another $2$ factor after applying $P$
invariance. Therefore in these processes we have $3^3=27$ independent
helicity amplitudes that we calculate using the Feynman rules shown in
eqs.(\ref{eq:GF3} and \ref{eq:GF4}). Here we can observe that the set of
the
chiral lagrangian parameters affecting the $W^{\pm}\gamma \rightarrow
W^{\pm}Z^0$ cross sections is the following: ${\alpha}_0$,
${\alpha}_1$, ${\alpha}_2$, ${\alpha}_3$, ${\alpha}_8$,
${\alpha}_9$, ${\alpha}_{11}$, ${\alpha}_{12}$ and ${\alpha}_{13}$.

In order to see the relative importance of the different channels we have
evaluated the contribution to the total cross section to obtain
$Z^0Z^0$ and $W^{\pm}Z^0$ pairs in $pp$ colliders from the different considered
subprocesses in a typical example. We have chosen the ${\alpha}_k$ parameters
that mimic the $MSM$ with a heavy scalar Higgs whose mass is
$m_H=1 TeV$ . The rates for  the contributions of the different
subprocesses to the total $Z^0Z^0$ final state,
$q\bar{q}$, $gg$, $Z^0Z^0$ and $W^+W^-$ are, respectively: $45$,
$17$, $3$, $35\%$ and to the total $W^{\pm}Z^0$ the rates are $61$,
$22$, $17\%$ coming from $q\bar{q'}$, $W^{\pm}Z^0$ and
$W^{\pm} \gamma$ fusion.

As it has been discussed above, the essential point in our calculations is that
we have not used the $ET$ \cite{ET,ETCL} to obtain the scattering
amplitudes of the gauge
bosons as it is customary. We have calculated the corresponding tree level
amplitudes for the gauge bosons and  projected them into their transversal or
longitudinal components. Therefore we do not have the limitations that
appear when the $ET$ and the chiral lagrangian formulation of the $ESB$
are used together \cite{ET}. The chiral lagrangian approach provides a
low-energy description of the $GB$ dynamics as an expansion on the
momenta over $4 \pi v$ and the $ET$ refers to the large energy relation between
$GB$ and the longitudinally polarized gauge bosons  $S$ matrix elements. In our
case we only need to fix an upper energy bound (as it was said $E_{\rm
max}=1.5 TeV$) so that we can safely apply $\chi PT$.

However, the $ET$  can be used as a helpful tool to test our
calculated longitudinally polarized amplitudes.
In order to check our results we have compared our tree level
expressions  with those calculated applying the $ET$ to the corresponding
 Goldstone boson amplitudes.
We have found both amplitudes to agree at lowest order in $g$ and $g'$
for the cases  $Z^0_LZ^0_L \rightarrow Z^0_LZ^0_L$,
$W^+_LW^-_L \rightarrow Z^0_LZ^0_L$ and $W^{\pm}_LZ^0_L \rightarrow
W^{\pm}_LZ^0_L$ as it was expected according to the results of
\cite{ETCL}. However, the details of this comparison will be described in
detail  elsewhere \cite{GLASGOW} since they concern  the applicability of the
$ET$ which has not been used in the present computation. Here we only wanted to
quote that our results are compatible with the information that one could
have obtained from the $ET$ on the above mentioned processes.

\section{Signatures and background}

As it was stressed in the previous section we have calculated the tree level
 scattering amplitudes, up to order $p^4$, corresponding to all the
helicity states of the gauge bosons, without using the $ET$.
This fact allows us to include the dependence on the ${\alpha}_k$
parameters in all the helicity  amplitudes. In contrast,
when the $ET$ is used, the only channel where the ${\alpha}_k$
coefficients are taken into account is the following:

\begin{displaymath}
V_LV_L  \rightarrow V_LV_L  \;\;\; (V=Z^0,W^{\pm})
\end{displaymath}

The definition of this channel as the signal was considered appropriate
because the
$V_LV_L  \rightarrow V_LV_L$ channels are expected to be strongly
interacting at high energies, if the $GB$ are, due to the $ET$.
 However, our computation makes it possible  for the first time to
study the effect of a strongly interacting ESB
 in other gauge boson
polarization states   by means of the ${\alpha}_k$
parameters. In fact, the aim of this work is to see how measurable
will these parameters be at the $LHC$. In order to make this point more
precise we must define the statistical significance corresponding to some
given value of the chiral parameters ${\alpha}_k$. With this goal in
mind we define our signal and our background as:

\begin{eqnarray}
({\rm Signal}) \;\;\;\;\;\; n_S       &  =  & |N(\{ {\alpha} \}) -
N(\{ {\alpha}^0 \})|
\label{eq:senal}
\end{eqnarray}
\begin{eqnarray}
({\rm Background}) \;\;\; n_B   &  =  &  N( \{ {\alpha}^0 \})
\label{eq:ruido}
\end{eqnarray}

where $N\{ {\alpha} \}$, $N\{ {\alpha}^0 \}$ are the total number of
$Z^0Z^0$ or $W^{\pm}Z^0$ pairs obtained for some given experimental cuts
when the
chiral parameters have been set to the values $\{ {\alpha} \}$ or $\{
{\alpha}^0 \}$. The background is defined in terms of some
reference model $\{
{\alpha}^0 \}$.
 For simplicity this model has been taken as the one with all the
parameters set to zero i.e. $ \alpha^0_k =0$.
We call it {\it Zero Model}  and,
incidentally, it
corresponds to the $MSM$ with an infinite Higgs mass.

For the final state $Z^0Z^0$ we have considered the following processes
that contribute to the signal and the background:

\begin{eqnarray*}
Z^0Z^0    \rightarrow       Z^0Z^0   \\
W^+W^-    \rightarrow       Z^0Z^0
\end{eqnarray*}

In addition we include the $Z^0Z^0$ production via
gluon fusion and $q\bar{q}$ annihilation which do not depend on
${\alpha}_k$ and therefore  contribute only to the  background.
The experimental signature for this process consists of four leptons as
the result
of the $Z^0Z^0$ decays. This gives a clean and distinct signal because
the $Z^0Z^0$ pairs can be fully reconstructed. The disadvantage is the
rather small leptonic branching ratio ($0.44\%$).

In the $W^{\pm}Z^0$ case we have considered the channels:

\begin{eqnarray*}
W^{\pm}Z^0       \rightarrow   W^{\pm}Z^0    \\
W^{\pm}\gamma    \rightarrow   W^{\pm}Z^0    \\
q\bar{q'}        \rightarrow   W^{\pm}Z^0
\end{eqnarray*}

We focus our attention on the gold-plated events where the $W^{\pm}$ and
$Z^0$ decay to the charged leptonic final states ($l=e,\mu,\nu$). The
corresponding branching ratio is $1.3\%$.
All  these cross sections depend on the chiral ${\alpha}_k$ parameters
so that they are taken into account in the signature and background
calculations.

The main source of $Z^0Z^0$ or $W^{\pm}Z^0$ pairs in $pp$
colliders is via quark-antiquark annihilation. The total production rate
of $gg \rightarrow Z^0Z^0$, in the studied cases, is $20-50\%$ than
that from
$q{\bar{q}} \rightarrow Z^0Z^0$ depending on the top quark mass (we
have chosen $m_t=170\,GeV$). On the other hand, the production rates of
$Z^0Z^0$ or
$W^{\pm}Z^0$, via gauge boson fusion are suppressed by powers
of (${\alpha}/ {\rm sin}^2 {\theta}_w$) due to the application of the
$W$ effective approximation \cite{fegamm,Daw} to obtain the initial
bosons from $pp$ beams, as we will see in next section.

\section{The proton and gauge boson structure functions}

Here we describe how to compute the total cross sections of the
different processes studied  to obtain $Z^0Z^0$ or $W^{\pm}Z^0$ pairs
in $pp$ colliders.

\begin{eqnarray*}
pp &  \rightarrow   &  (q\bar{q} \rightarrow V_3V_4) + X    \\
pp &  \rightarrow   &  (gg \rightarrow Z^0Z^0) + X    \\
pp &  \rightarrow   &  (V_1V_2  \rightarrow  V_3V_4) + X
\end{eqnarray*}

We have to integrate the differential cross section for the subprocess,
$(d{\sigma}/d {\rm cos} {\theta})$, with the distribution functions of
the
quark, antiquark and gluon (given by $f_i$, $f_j$ and $g$) inside the
proton:

\begin{eqnarray}
\sigma (pp \rightarrow (q\bar{q'} \rightarrow V_3V_4) + X) & = &
\sum_{i,j} \int\int\int dx_1dx_2 dcos\theta f_i(x_1,Q^2)f_j(x_2,Q^2)
\nonumber  \\
							   &   &
\frac{d \hat{\sigma}}{d cos{\theta}}(q\bar{q'} \rightarrow V_3V_4)
\nonumber  \\
\sigma (pp \rightarrow (gg \rightarrow Z^0Z^0) + X)        & = &
\int\int\int dx_1dx_2 dcos\theta g(x_1,Q^2)g(x_2,Q^2)
\nonumber  \\
							   &   &
\frac{d \hat{\sigma}}{d cos{\theta}} (gg \rightarrow Z^0Z^0)
\label{eq:secc}
\end{eqnarray}

These formulae are used to compute the processes:

\begin{eqnarray*}
q\bar{q}  & \rightarrow   &  Z^0Z^0      \\
q\bar{q'} & \rightarrow   &  W^{\pm}Z^0  \\
gg        & \rightarrow   &  Z^0Z^0
\end{eqnarray*}

In order to compute the number of events of $Z^0Z^0$ and $W^{\pm}Z^0$
produced
in $pp$ collisions via gauge boson fusion we apply the effective $W$
approximation \cite{fegamm,Daw} and we use the formula:

\begin{eqnarray}
\sigma (pp \rightarrow (V_1V_2 \rightarrow V_3V_4) + X)   &  =  &
\sum_{i,j} \int\int dx_1dx_2 dcos\theta f_i(x_1,Q^2)f_j(x_2,Q^2)
\nonumber    \\
 &     &  \int\int d\hat{\tau} d\hat{\eta}  \frac{{\partial}^2
L}{{\partial}\hat{\tau}{\partial}\hat{\eta}}
\frac{d \hat{\sigma}}{d cos{\theta}} (V_1V_2 \rightarrow V_3V_4)
\label{eq:secc2}
\end{eqnarray}

Thus the total cross section in $pp$ colliders (like $LHC$) can be
written as the result of the convolution of the subprocess cross section
with the $V_1V_2$ pair luminosity  in $pp$ beams. This luminosity
is calculated from the convolution of the  double bremsstrahlung of the
$(V_1V_2)$ from the quark structure functions. Thus,
$\partial^2 L/\partial\hat{\tau}\partial\hat{\eta}$
is the luminosity function for the gauge boson pair $V_1^h V_2^h$
to be radiated from the quark pair $q_iq_j$. It depends on the
helicity state, transversal or longitudinal, of the initial bosons $V_1$
and $V_2$. Therefore  we have to separate in our computations the
contribution of the
different polarization channels of the initial bosons.

\begin{eqnarray*}
Z^0Z^0         &   \rightarrow  &  Z^0Z^0       \\
W^+W^-         &   \rightarrow  &  Z^0Z^0       \\
W^{\pm}Z^0     &   \rightarrow  &  W^{\pm}Z^0   \\
W^{\pm}\gamma  &   \rightarrow  &  W^{\pm}Z^0
\end{eqnarray*}

The amplitudes and differential cross sections for all these
processes have been obtained as it was described in the previous
sections. According to the effective $W$ approach we take the following
functions corresponding to the probability of a gauge boson $V_T$ or $V_L$
($V=W^{\pm}, Z^0$) or a photon, to be radiated from the quark $q$,
with a momentum fraction of the quark, $x$.

\begin{eqnarray}
f_{q/V^T}(x)  &  = &   f_V \frac{x^2+2(1-x)}{2x} {\rm ln}
\left( \frac{E^2}{M^2_V} \right) \;\;\;\;\; V=W,Z      \nonumber  \\
f_{q/V^L}(x)  &  =  &   f_V \frac{1-x}{x}  \nonumber  \\
f_{q/ \gamma}(x)   &  =  & \frac{\alpha}{2\pi} {\epsilon}^2_q
\frac{1+(1-x)^2}{x} {\rm ln} \left( \frac{E^2}{m^2_q} \right)
\label{eq:Wef}
\end{eqnarray}

where the values of the $f_{{V_i}'s}$ depend on the particular gauge
boson as well as on the type of quark that it comes from

\begin{eqnarray*}
f_W(x)  &  =  &  \frac{\alpha}{4 \pi  x_W}  \\
f_{Z_{u\bar{u}}}(x)   &  =  & \frac{\alpha}{16 \pi x_W (1-x_W)}
\left[ 1+(1-\frac{8}{3} x_W)^2 \right]    \\
f_{Z_{d\bar{d}}}(x)   &  =  & \frac{\alpha}{16 \pi x_W (1-x_W)}
\left[ 1+(1-\frac{4}{3} x_W)^2 \right]
\end{eqnarray*}

where $x_W={\rm sin}^2 {\theta}_W$.

The variables $\tau$ and $\eta$ are related to the momentum fractions of
the quarks by $x_{1,2}=\sqrt{\tau} e^{{\pm}\eta}$.
The connection between the variables $\hat{\tau}$, $\hat{\eta}$ and the
momentum fractions of $V_1$, $V_2$ respect to $q_i$, $q_j$, $\hat{x_1}$
and $\hat{x_2}$ is given by
$\hat{x_{1,2}}=\sqrt{\hat{\tau}}e^{\pm {\hat{\eta}}}$.
Note that the rate of transversally polarized gauge bosons obtained is enhanced
by the logarithmic factors eq.(\ref{eq:Wef}) with respect to the
longitudinal gauge boson production.

The structure functions of quarks and gluons we use are those of $EHLQ$
\cite{EHLQ}, set $II$  with $\Lambda = 290 MeV$ and we neglect the
contribution of
the top quark to the proton sea. However we have also studied the
effect of changing the structure function on our results (see below).
The assignment for the $Q^2$ appearing in the distribution functions is $Q^2 =
\hat{s}$ for $q\bar{q'}$ and $gg$ processes, and $Q^2=M^2_W$ for elastic gauge
bosons scattering processes.  The resulting integrals are computed using the
$VEGAS$
 Monte Carlo
program \cite{Vegas}.

With the described machinery we have built up a big code which
computes the total number of expected gold-plated events
(for some given $LHC$ integrated luminosity) in terms  of the
chiral parameters $\alpha_k$ and the final
state cuts .

Initially we impose the following cuts on the
invariant
mass of the weak boson pair, the transverse momentum of the final $Z^0$
boson and the rapidities of both bosons:

\begin{displaymath}
\begin{array}{lclll}
200 GeV            &  \leq  &  \sqrt{\hat{s}} &  \leq  &
1500 GeV  \nonumber   \\
{p_T}_Z            &  \geq  &     10 GeV      &        &
\nonumber     \\
| y_{1_{\rm max}} | &  =  & | y_{2_{\rm max}} | &   =  &  2.5
\end{array}
\end{displaymath}

{}From now on, we will call them the minimal cuts. The upper limit in
the invariant mass ensures the validity of $\chi PT$ approach. For the rest of
the paper we compute the number of $Z^0Z^0$ and $W^{\pm}Z^0$
obtained in the $LHC$ with an integrated luminosity of $3 \times 10^5
pb^{-1}$. This approximately corresponds to a total $LHC$ working time
of 1 year
($3\times 10^7 sec$) assuming a luminosity
${\cal L} = 10^{34} cm^{-2}s^{-1}$.

\section{Parameter sensitivity, optimal cuts and measurable parameters}

In previous sections we have presented in detail how we have
computed the number
of $Z^0Z^0$ or  $W^{\pm}Z^0$ pairs obtained at the $LHC$.
If we fix the integrated luminosity, the $pp$ center of mass frame
energy for the $LHC$ ($\sqrt{s}=16TeV$), and the upper bound on the
invariant mass at the subprocesses ($\sqrt{\hat{s}_{{\rm max}}}=1.5
TeV$), the total cross sections will depend on the chiral ${\alpha}_k$
parameters and on some kinematical cuts. For the sake of simplicity we
will only study the subset of the ${\alpha}_k$ coefficients which
is needed to reabsorb the one-loop divergences obtained  calculating with
the chiral Lagrangian to lowest order. They are ${\alpha}_0$,
${\alpha}_1$,
${\alpha}_2$, ${\alpha}_3$, ${\alpha}_4$ and ${\alpha}_5$ \cite{Long}
(however, our code includes as well the
contribution of the other
chiral parameters). There is also the possibility to modify
our results choosing different kinematical cuts,
$c_l$ ($c_1 = \sqrt{\hat{s}_{\rm min}}$, $c_2 = {p_{TZ}}_{\rm
min}$ and $c_3 = y_{1_{\rm max}}=y_{2_{\rm max}}$). Thus we can define
two vectors:
${\alpha}$ and $c$, as follows:
\begin{eqnarray}
{\alpha} & = & ({\alpha}_0,{\alpha}_1,{\alpha}_2,{\alpha}_3,
{\alpha}_4,{\alpha}_5)       \nonumber     \\
c  & = & (c_1,c_2,c_3)
\label{eq:vectors}
\end{eqnarray}
which are inputs of our code; whereas the
number of $Z^0Z^0$ ($i=1$) or $W^{\pm}Z^0$ ($i=2$) events is the output.
Therefore we write   the result of our computations as $N^{(i)}({\alpha};c)$.
In principle one could expand these functions around the {\it Zero
Model} (we defined it previously as $\{ \alpha^0_k=0 , \forall k \}$)
and the minimal cuts defined above so that one could
write:

\begin{eqnarray}
N^{(i)} ({\alpha}; c)  &  =  &   N^{(i)} ({\alpha}^0; c^{\rm min}) +
\sum^{5}_{k=0} \frac{\partial N^{(i)}}{\partial {\alpha}_k}
{\mid}_{{\alpha}_k=0} {\alpha}_k + \sum^{3}_{l=1}
\frac{\partial N^{(i)}}{\partial c_l}{\mid}_{c_l=c^{\rm min}_l}
(c_l - c^{\rm min}_l)  \nonumber   \\
 &  & + O({\alpha}^2) +   O({(c-c^{\rm min})}^2)
\end{eqnarray}

If the dependence of the number of events on the parameters and the
cuts  were approximately linear, this formula could be used to compute
$N^{(i)}({\alpha};c)$. However, we will see that this is not always the case.
Indeed, $N^{(i)}({\alpha};c)$ is a
polynomial in $\alpha_k$ and nonlinear terms  can
become important even for moderate values of these parameters.

In order to see how the number of events changes when one of the
${\alpha}_k$ parameters varies by some amount $\Delta {\alpha}_k$,
we fix the kinematical bounds $c_l$ to
the minimal cuts and set the other chiral parameters to zero.
We choose the following set of values
for $\Delta {\alpha}_k$:

\begin{displaymath}
\Delta {\alpha}_k = \pm 10^{-3}, \pm 5 \times 10^{-3}, \pm 10^{-2}
\end{displaymath}

The result of our computations can be found in Figures \ref{fig:sena0},
to \ref{fig:sena5}.
We can also define the sensitivity function $s^{(i)}_k(c)$ associated to
the ${\alpha}_k$ parameter, with the kinematical cuts $c$, and for the
final state $i$, as follows:

\begin{equation}
s^{(i)}_k (c) \equiv \frac{\partial N^{(i)}
({\alpha};c)}{\partial {\alpha}_k} {\mid}_{{\alpha}_k=0}
\label{eq:sens}
\end{equation}

The different sensitivity functions for the minimal cuts that we have
obtained
are displayed in Table \ref{tab:line}. Obviously, the $|s^{(i)}_k
(c)|$  values are a measure of the variations of the
number of events with ${\alpha}_k$. However, they are not a direct measure
of the statistical significance of the corresponding parameter variation.

We can also study the number of $Z^0Z^0$ and $W^{\pm}Z^0$
event
distributions with respect to the cuts given in the minimal invariant mass
($c_1$), the minimal transverse momentum of $Z^0$ in the final state ($c_2$)
and in the maximal rapidity of the final bosons ($c_3$). In these
computations we have fixed the ${\alpha}$ parameters at their values
in the {\it Zero Model}, and the other two kinematical cuts have been
set to their minimal values. The results obtained in this way are shown
in Figures \ref{fig:mvcut} to \ref{fig:yicut}.

In  order to perform a statistical analysis, we
define the $r_k$ function as:

\begin{equation}
\begin{array}{lclcl}
r^{(i)}_k (c) & = & \frac{| N^{(i)}({\alpha}_k; c)
- N^{(i)}({\alpha}^0; c) |}{\sqrt{N^{(i)}({\alpha}^0;c)}}
 & = & \frac{n_S}{\sqrt{n_B}}
\end{array}
\label{eq:rk}
\end{equation}

The $r^{(i)}_k$ function is a measure of the statistical significance of
the signal (corresponding to the increments $\Delta {\alpha}_k$)
relative to the background, provided $N^{(i)}({\alpha}_k; c)$ is large
enough to apply the Central Limit Theorem. In this case $r_k$ is an
estimate of the number of sigmas, and therefore it defines the
confidence level
for the hypothesis that ${\alpha}_k$ is different from zero. In
order to evaluate $r^{(i)}_k$ we use the number of events that was
previously
obtained. The resulting $r^{(i)}_k$ functions are shown in Figures
\ref{fig:siga0} to \ref{fig:siga5}.

{}From Figures  \ref{fig:mvcut} and \ref{fig:ptcut},
it is clear that there is no linear dependence of
$N^{(i)}({\alpha}; c)$ on  $c_1$ and $c_2$. Moreover, looking at Figures
\ref{fig:sena0} at \ref{fig:sena5} and from Figures
\ref{fig:siga0} at \ref{fig:siga5}, nonlinear behaviour
is observed with respect to ${\alpha}_1$, ${\alpha}_2$ and
${\alpha}_3$ (for $W^{\pm}Z^0$ final state) nor when $\mid
{\alpha}_4 \mid$ or $\mid {\alpha}_5 \mid$ are bigger than $0.001$  in
both channels, $Z^0Z^0$ and $W^{\pm}Z^0$.

By looking at Figures \ref{fig:siga0} to \ref{fig:siga5}, we can see
that the slope corresponding to
the different curves $r^{(i)}_k(c^{\rm min})$ varies according to the
different values of ${\alpha}_k$ and the channel ($Z^0Z^0$ or
$W^{\pm}Z^0$). Indeed, the greater is the  slope, the higher is the
statistical significance. Moreover it is already possible
to observe that some of the ${\alpha}$
parameters have no chances to be measured at the
$LHC$, at least in the way described here. For this reason, in the following
we
will concentrate our statistical analysis on the set of  ${\alpha}$ parameters
which
can be considered as {\it potentially measurable}. Being more
precise, we define a parameter $\alpha_k$
as   {\it potentially
measurable} in a given channel when  the $r^{(i)}_k(c^{\rm min})$
corresponding to the maximal considered variation
($\mid \Delta {\alpha}_k \mid = 0.01$), is bigger than  $0.5$.
According to this we find that the
{\it potentially measurable}  ${\alpha}_k$ are the following:

\begin{itemize}

\item ${\alpha}_3<0$ and ${\alpha}_3>0$ (both in $Z^0Z^0$ and
$W^{\pm}Z^0$ channels)

\item ${\alpha}_4<0$ ($W^{\pm}Z^0$) and ${\alpha}_4>0$ ($Z^0Z^0$)

\item ${\alpha}_5<0$ ($W^{\pm}Z^0$) and ${\alpha}_5>0$ ($Z^0Z^0$)

\end{itemize}

In the following we carry out an statistical analysis with all these
{\it potentially measurable}
parameters in the corresponding final state. To do so, the next
step is to
 look
for the optimal cuts in the minimal invariant mass and transverse
momentum for the detection of a certain  $\Delta
{\alpha}_k = ({\alpha}_k-{\alpha}^0_k)$ different from zero.
Initially we impose the minimal cuts on the kinematical variables.
The optimization procedure we use is to search for the pair of
cuts ($\sqrt{\hat{s}}$, $p^c_{TZ}$) in the kinematical region

\begin{eqnarray}
200 GeV \;\; \leq & \sqrt{\hat{s}} & \leq \;\; 1500 GeV    \nonumber  \\
10 GeV  \;\; \leq & p_{TZ} & \leq \;\; \sqrt{\hat{s}_{\rm max}/4-m^2_z}
\label{eq:lig}
\end{eqnarray}

that maximize
the function $r^{(i)}_k({\alpha}_k; \sqrt{\hat{s}^c}, p^c_{TZ})$ defined
as  follows:

\begin{equation}
r^{(i)}_k ({\alpha}_k; \sqrt{\hat{s}}^c, p^c_{TZ}) = \frac{ | N^{(i)}
({\alpha}_k; \sqrt{\hat{s}} > \sqrt{\hat{s}^c}, p_{TZ}>p^c_{TZ}) -
N^{(i)} ({\alpha}^0_k; \sqrt{\hat{s}} > \sqrt{\hat{s}^c},
p_{TZ}>p^c_{TZ}) |}
{\sqrt{N^{(i)} ({\alpha}^0_k; \sqrt{\hat{s}} > \sqrt{\hat{s}^c},
p_T>p^c_T)}}
\label{eq:opti}
\end{equation}

To find the optimal cuts for the different parameters and channels we
build a
bidimensional grid on the plane ($\sqrt{\hat{s}},p_{TZ}$) so that the
different
points are separated from each other by the increments ($\Delta
\sqrt{\hat{s}} =
50 GeV$, $\Delta p_T = 50 GeV$). For each pair of points in this grid
($\sqrt{\hat{s}^c}$, $p^c_{TZ}$) we compute the total number of events
in the
{\it Zero Model} and in a $ESB$ scenario corresponding to  a
certain positive or negative $\Delta{\alpha}_k$. Finally, we
evaluate the $r^{(i)}_k ({\alpha}_k; \sqrt{\hat{s}^c}, p^c_{TZ})$
function. The optimal cuts, $c^{\rm op}$, are those which maximize
eq.(\ref{eq:opti}), and therefore,
the confidence level. Of course, these $c^{\rm op}$ which have been
found in this way depend slightly on the particular choice of
$\Delta{\alpha}_k$ for each ${\alpha}_k$ parameter. In our computation
we have taken the typical values:

\begin{equation}
\Delta {\alpha}_k = {\alpha}_k - {\alpha}^0_k = {\alpha}_k = \pm 0.005
\end{equation}

 The results are collected in Tables \ref{tab:opti1} and
\ref{tab:opti2}. There it
is displayed, for each $\Delta{\alpha}_k$ in the corresponding channel,
the optimal cuts, the number
of events that satisfies these cuts and the statistical significance
function $r_k$ obtained with the minimal and the optimal cuts. In all
cases we have fixed a maximal rapidity of $2.5$.

The $c^{\rm op}$ we have obtained by maximizing the
$r_k^{(i)}$ function  eq.(\ref{eq:opti})
will be considered from now on as the optimal cuts to detect
either a $\Delta{\alpha}_k < 0$ or a $\Delta{\alpha}_k > 0$.
For example,  if we wanted to
observe the signature corresponding to a heavy Higgs $SM$-like
$ESB$ scenario,
the best choice would be the $Z^0Z^0$ final
state with the optimal cuts obtained for $\Delta{\alpha}_5>0$
($\sqrt{\hat{s}_{\rm min}} = 1150 GeV$, $p_{TZ{\rm min}} = 400 GeV$). This is
so because, at the  tree level order, the only ${\alpha}_k$ different from
zero needed to mimic this scenario is ${\alpha}_5$ (${\alpha}_5=v^2/(8m^2_H)$).

As it can be seen in Table \ref{tab:opti2},
the optimization procedure has clearly improved the statistical
significance $r_k$ in most of the studied cases. Only when we tried
to optimize the signature corresponding to a ${\alpha}_3>0$ and
${\alpha}_3<0$, both in the $Z^0Z^0$ final state, we obtained no
significant  improvement
with respect to the minimal cuts. Therefore
 we find it impossible to detect a
signature of a negative or positive ${\alpha}_3$ in $Z^0Z^0$ events.

Finally, we have carried out a similar optimization procedure
for $\alpha_4$ and $\alpha_5$ in which
we searched for a slightly different pair of optimal cuts:
($\sqrt{\hat{s}_{\rm max}}$, ${p_{TZ}}_{\rm min}$).
We obtained $c^{\rm op}$ = ($1500 GeV$, $500 GeV$) for
${\alpha}_4=-0.005$ ($W^{\pm}Z^0$), ${\alpha}_4=0.005$ ($Z^0Z^0$) and
${\alpha}_5=0.005$ ($Z^0Z^0$) and the $r_k$ function was respectively
$6.02$, $1.56$ and $3.82$. On the other hand, we also found the optimal
cuts to detect ${\alpha}_5=-0.005$ in the $W^{\pm}Z^0$ final state
($c^{\rm op}=(1500 GeV, 400 GeV)$) and the reached statistical
significance was $2.97$. These results are at most equal to those of
Table \ref{tab:opti2}.

Moreover,
we have repeated our whole study to find other sets of optimal cuts
belonging to the parameter space, such as ($\sqrt{\hat{s}_{\rm max}}$,
${p_{TZ}}_{\rm min}$), ($\sqrt{\hat{s}_{\rm max}}$, ${p_{TZ}}_{\rm
max}$),
($\sqrt{\hat{s}_{\rm min}}$, ${p_{TZ}}_{\rm max}$). In all these cases,
the statistical significances obtained are smaller than those
corresponding to the first choice of $c^{\rm op}$ ($c^{\rm
op}=(\sqrt{\hat{s}_{\rm min}}, {p_{TZ}}_{\rm min}$)). Therefore, in the
following we will apply this type of optimal cuts.

As it was mentioned in the introduction the main aim of this work is to
determine which
chiral Lagrangian
parameters will be more easily measurable at the $LHC$ and the size of their
corresponding statistical errors.
Initially, we made a criterion to choose a set of {\it potentially
measurable} ${\alpha}_k$ parameters, on which we have concentrated our
analysis. After carrying out the described optimization procedure,
we conclude that the statistical significance to measure most of these
chiral Lagrangian coefficients, has clearly increased.
In the following we will treat only these {\it measurable} ${\alpha}_k$.
The set of the ${\alpha}_k$ parameters obtained as the
{\it measurable} ones are contained in Tables \ref{tab:opti1} and
\ref{tab:opti2}. Here we exclude the
study of ${\alpha}_3$ in the $Z^0Z^0$ final state, as it was previously
argued.

In order to obtain an estimation of the statistical errors ${(\Delta
{\alpha}_k)}_{\rm stat.}$ we will work under some hypothesis. We will assume
a linear behaviour in ${\alpha}_k$ and that the
number of events is  large enough to consider  $N^{(i)}(\alpha;c)$  following
a gaussian distribution.

At this point, we define the statistical error ${(\Delta
{\alpha}_k)}_{\rm stat.}$ associated to a certain ${\alpha}_k$ as the
minimal value of this parameter that could be detected in the $LHC$ with
a statistical significance, eq.(\ref{eq:rk}), equal to one
sigma, after applying its corresponding optimal cuts.  Using the results
displayed in Table \ref{tab:opti2} which satisfy the optimal cuts, we
obtain the
${(\Delta {\alpha}_k)}_{\rm stat.}$  in this way, and we show the
results in Table \ref{tab:error}. As it could be expected, the smallest
${(\Delta {\alpha}_k)}_{\rm stat.}$ value corresponds to the biggest
statistical significance, which is
${(\Delta {\alpha}_4)}_{\rm stat.}=8.31 \times 10^{-4}$, in the
$W^{\pm}Z^0$ final state.

\section{The effect of the structure function indetermination}

In all  the previous calculations we have used the $EHLQ$
structure functions \cite{EHLQ} (set $II$ with $\Lambda=290 MeV$) to obtain the
total number of $Z^0Z^0$ and $W^{\pm}Z^0$ events. Here we wonder how our
results could be affected by the indetermination in the parton distribution
 functions. In
order to estimate this effect  we have selected two new
parametrizations of the structure function: the set $MRSD-$ (Martin,
Roberts and Stirling \cite{MRS}) and the $GRVHO$ (Gl\"{u}ck, Reya and
Vogt \cite{GRV}). Both them exhibit
a similar behaviour of sea quarks and gluon distributions at low $x$,
that grow when $x \rightarrow 0$.
The values of $\Lambda$ considered in these two sets are the following:
${\Lambda}_{\rm MRS}=230 MeV$, ${\Lambda}_{\rm GRV}=200 MeV$.
 With these new
structure functions we evaluate the total number of $Z^0Z^0$ or
$W^{\pm}Z^0$ events, in the {\it Zero Model}  with the different optimal cuts.
The way we proceed is to compare our estimation of
the number of standard deviations obtained by changing these parton
density
function sets (we will call it $r^{(i)}_{\rm struc.}$), with the
statistical significance corresponding to a
certain signal, after having applied the optimal cuts
(the $r^{(i)}_k(c^{\rm op}$) functions  shown in Table
\ref{tab:opti2}).
We define $r^{(i)}_{\rm struc.}$ similarly as in
eq.(\ref{eq:rk}):

\begin{equation}
r^{(i)}_{\rm struc.}(c)  = \frac{| N_{set}^{(i)}({\alpha}^0; c)
- N_{set'}^{(i)}({\alpha}^0; c) |}
{\sqrt{N_{set}^{(i)}({\alpha}^0;c)}}
\label{eq:rstruc}
\end{equation}

Now, to obtain  $r^{(i)}_{\rm struc.}$ we need to evaluate
the total number of events in the {\it Zero
Model}, with certain kinematical cuts, $c$, but using two different
parametrizations called {\it set} and {\it set'} for the structure
functions.

In Table \ref{tab:festru} we show the results of our calculations.
We represent
the total number of $Z^0Z^0$ and $W^{\pm}Z^0$ events obtained for
the optimal cuts, when the
$EHLQ$ (set II), $MRSD-$ and $GRVHO$ sets are used, as well as the number of
sigmas $r_{\rm struc.}$ eq.(\ref{eq:rstruc})
corresponding to change one of these sets with
respect to the others:

\begin{itemize}

\item $r_{\rm struc.} \equiv r_{12}$ ($EHLQ$ with
respect to $MRSD-$)

\item $r_{\rm struc.} \equiv r_{13}$ ($EHLQ$ with respect to $GRVHO$)

\item $r_{\rm struc.} \equiv r_{23}$ ($MRSD-$ with respect to $GRVHO$).

\end{itemize}

Now, we pay attention to the $r_k$ functions in Table \ref{tab:opti2}
and
$r_{\rm struc.}$ in Table \ref{tab:festru} and we compair them.
We will say that the typical chosen value of ${\alpha}_k$
(${\alpha}_k=-0.005$ or ${\alpha}_k=0.005$) can be measured in the $LHC$
with a statistical significance given by $r_k$, if the following
inequality is verified:

\begin{equation}
r_k >> r_{\rm struc.}
\label{eq:ineq}
\end{equation}

When we make $r_{\rm struc.}=r_{12}$ or $r_{\rm struc.}=r_{13}$, we must
reject the value ${\alpha}_3=0.005$ in $Z^0Z^0$ and $W^{\pm}Z^0$
channels, and
${\alpha}_3=-0.005$ in $Z^0Z^0$ final state, because  relation
(\ref{eq:ineq}) is violated.
However, if we take $r_{\rm struc.}=r_{23}$ we can measure positive and
negative values of ${\alpha}_3$, ${\alpha}_4$ and ${\alpha}_5$
parameters in some $Z^0Z^0$ or $W^{\pm}Z^0$ final state, with a
statistical significance given by $r_k$. On the opposite, for
${\alpha}_3=\pm 0.005$ in $Z^0Z^0$ final state, eq.(\ref{eq:ineq}) is
not
fulfilled, so we can only measure these values of ${\alpha}_3$ through the
$W^{\pm}Z^0$ channel. This last choice ($r_{\rm struc.}=r_{23}$) is most
appropriate since $MRSD-$ and  $GRVHO$ agree much better with some
recent experimental results \cite{Terron} obtained at $HERA$. In any case, it
can be expected that future new experimental data coming from $HERA$ and even
from the $LHC$ itself, will allow us to reduce the size of these uncertainties
$r_{\rm struc.}$.

\section{Renormalization effects}

The other effect we are going to take into account is the
dependence of the ${\alpha}_k$ parameters and the couplings $g$ and
$g'$, on the energy. Our whole previous study was carried out at tree
level order. Now we want to include some quantum effects by means of the
Renormalization Group Equations ($RGE$). By that we will understand the one
loop contributions coming from ${\cal L}^{(2)}$ and ${\cal L}_{\rm YM}$ plus
the tree level contributions coming from ${\cal L}'_1 + {\cal L}^{(4)}$ to the
corresponding beta functions.

As it is well known from the $RGE$, the  renormalized Green function with the
renormalized parameters ${\lambda}$ and $m$ at the
renormalization scale $\mu$ and the same Green function with the parameters at
another scale $\mu e^t$,  are related by the equation:

\begin{equation}
G^{(n)}_R(p_i,\lambda,\mu) = G^{(n)}_R(p_i,\bar{\lambda}(t),
\mu e^t) {\rm exp} [-n \int^t_0 \gamma(\bar{\lambda}(t')) dt']
\label{eq:EGR}
\end{equation}

Where we include in ${\lambda}$ all  the
${\alpha}_k$ parameters and the electroweak coupling constants, $g$ and $g'$.
Thus the renormalized parameters  are
scale dependent. The $\bar{\lambda}$ stands for all the running coupling
constants that depend on the renormalization scale. In order to obtain these
$\bar{\lambda}$ we need to calculate the $\beta({\lambda})$ and
$\gamma({\lambda})$
functions, following the well known  $RGE$ techniques:

\begin{eqnarray}
\beta (\bar{\lambda}) & = & \frac{\partial \bar{\lambda}(\lambda,t)}
{\partial t}     \nonumber  \\
\gamma (\bar{\lambda}(t)) & = & \frac{1}{2} \frac{\partial {\rm log}
Z_{3V}}{\partial t}
\label{eq:begam}
\end{eqnarray}

At present, when using the effective theory to describe the $ESB$ sector
we only know
these functions to one loop-order. The exponential factor in
eq.(\ref{eq:EGR})
is the anomalous dimension term, that depends on the wave function
renormalization. We have estimated the order of magnitude of this effect
to calculate the statistical significance $r_k$, in a typical case, with
a subprocess energy of $1TeV$. We obtained changes on $r_k$ of $0.1\%$.
Therefore they are completely irrelevant. Thus we
neglect this contribution and we will only consider the dependence on the
renormalization scale, $\mu$,
of ${\alpha}_k$, $g$ and $g'$, in  eq.(\ref{eq:EGR}). We refer
to this approach as the tree level approximation improved by the $RGE$.

Now, we have to calculate the running coupling constants
${\alpha}_k(\mu)$,  $g(\mu)$ and $g'(\mu)$, and replace their tree
level values in our cross section formulae by the corresponding energy
dependent running coupling constants evaluated at the subprocess center of
mass energy (we have taken $\mu=\sqrt{\hat{s}}$).

In order to do so we begin with the ${\alpha}_k$ parameters and we use
the dimensional
regularization scheme since it is the most appropriate for gauge
theories as well as for
non-linear sigma models. In \cite{Long}, Longhitano obtained all the
divergences
that appear in one-loop
diagrams calculated with the effective lagrangian at lowest order
(${\cal L}^{(2)}+{\cal L}_{\rm YM}$). They could be absorbed by
redefinitions of a subset of the ${\alpha}_k$ parameters
($k=0,1,2,3,4,5$). By
means of the dependence of ${\alpha}_k$ on the $\epsilon$ parameter
\cite{Long}, it is easy to
obtain the following ${\alpha}_k$ running  expressions:

\begin{eqnarray}
{\alpha}^R_0(\mu) & = & {\alpha}^R_0({\mu}_0) - \frac{1}{{16 \pi}^2}
\frac{3}{4} {\rm tg}^2 {\theta}_{\rm w} {\rm
log} \left( \frac{{\mu}_0}{\mu} \right)
\nonumber   \\
{\alpha}^R_1(\mu) & = & {\alpha}^R_1({\mu}_0) - \frac{1}{{16 \pi}^2}
\frac{1}{6} {\rm tg}{\theta}_{\rm w} {\rm log}
\left( \frac{{\mu}_0}{\mu} \right)
\nonumber   \\
{\alpha}^R_2(\mu) & = & {\alpha}^R_2({\mu}_0) - \frac{1}{{16 \pi}^2}
\frac{1}{12} {\rm tg}{\theta}_{\rm w} {\rm log}
\left( \frac{{\mu}_0}{\mu} \right)
\nonumber   \\
{\alpha}^R_3(\mu) & = & {\alpha}^R_3({\mu}_0) - \frac{1}{{16 \pi}^2}
\frac{1}{12} {\rm log} \left( \frac{{\mu}_0}{\mu} \right)   \nonumber
\\
{\alpha}^R_4(\mu) & = & {\alpha}^R_4({\mu}_0) + \frac{1}{{16 \pi}^2}
\frac{1}{6} {\rm log} \left( \frac{{\mu}_0}{\mu} \right)   \nonumber
\\
{\alpha}^R_5(\mu) & = & {\alpha}^R_5({\mu}_0) + \frac{1}{{16 \pi}^2}
\frac{1}{12} {\rm log} \left( \frac{{\mu}_0}{\mu} \right)
\label{eq:arun}
\end{eqnarray}

Apart from these equations, we need also to calculate the running
of the gauge couplings $g_R({\mu})$ and $g'_R({\mu})$. As usual we have

\begin{eqnarray*}
g_R & = & Z_3^{3/2} Z_1^{-1} g_0    \\
g'_R & = & {Z'}_3^{1/2} g'_0
\end{eqnarray*}

The only differences with respect to the linear model, given by $\delta
Z_3$ and $\delta Z'_3$ functions,  come from the scalar sector
contribution. In our case, they are

\begin{eqnarray*}
\delta Z_3  & = & \frac{g^2}{16 {\pi}^2} \frac{1}{6 \epsilon} \\
\delta Z'_3 & = & \frac{g'^2}{16 {\pi}^2} \frac{1}{6 \epsilon}
\end{eqnarray*}

Taking into account loops of scalars, fermions, ghosts and gauge bosons
coming from the rest of the $SM$ particles we arrive to the following
values of the $\beta$ functions eq.(\ref{eq:begam}), ${\beta}_g$ and
${\beta}_{g'}$:

\begin{eqnarray}
{\beta}_g & = & - C_g g^3   \nonumber    \\
{\beta}_{g'}  & = & - C_{g'} {g'}^3
\label{eq:betas}
\end{eqnarray}

where

\begin{eqnarray*}
C_g     & = & \frac{1}{16 {\pi}^2} \frac{13}{4}   \\
C_{g'}  & = & - \frac{1}{16 {\pi}^2} \frac{27}{4}
\end{eqnarray*}

The eqs.(\ref{eq:betas}) are standard corresponding for
gauge theories. The first one leads to asymptotic freedom since $C_g>0$,
whereas the second corresponds to an abelian gauge
theory. Therefore one has:

\begin{eqnarray}
g_R^2(\mu) & = & \frac {  g_R^2(\mu_0)  } {1 + \frac{1}{16 \pi^2}
\frac{13}{2} g^2_R(\mu_0) {\rm log} \frac{\mu}{\mu_0}    }  \nonumber  \\
{g'}_R^2(\mu) & = & \frac{  {g'}_R^2(\mu_0)  } {1 - \frac{1}{16
\pi^2}
\frac{27}{2} {g'}^2_R(\mu_0) {\rm log} \frac{\mu}{\mu_0}   }
\label{eq:gsR}
\end{eqnarray}

Once we have obtained the expression giving the dependence of the ${\alpha}_k$,
$g$ and $g'$ on the renormalization scale eq.(\ref{eq:gsR}), we
can study cuantitatively this effect. The way we proceed  is the
following:

\begin{itemize}

\item[-] First we factorize the $g$ and $g'$ coupling constants so that
the cross
sections appear with the same power in the electroweak couplings than
in the $MSM$. We are referring only
to the dominant terms. Moreover, as we can see in the Appendix,
eqs.(\ref{eq:GF2}, \ref{eq:GF3} and \ref{eq:GF4}), higher powers
on these
coupling constants appear as factors of the ${\alpha}_k$ parameters.

\item[-] In all  our calculations, for the sake of simplicity, we are
considering the physical masses \cite{datos}:
$m_W=80.6 GeV$, $m_Z= 91.1GeV$ and $m_q=0$ (except to in $gg \rightarrow
Z^0Z^0$ process where we have taken a top quark in the loop with a mass
of $170 GeV$).

\item[-] We replace in the cross section  the constant values of
${\alpha}_k$, $g$ and $g'$ by those which are renormalized and scale dependent,
${\alpha}^R_k(\mu)$, $g_R(\mu)$ and $g'_R(\mu)$ given by
eqs.(\ref{eq:arun} and \ref{eq:gsR}). We take
$\mu=\sqrt{\hat{s}}$, where $\sqrt{\hat{s}}$ is the invariant mass (the
center of mass energy in the considered subprocess) and ${\mu}_0=m_Z$.
Thus  we have substituted the tree level values of ${\alpha}_k$, $g$ and
$g'$ by the running coupling constants depending on the energy scale of
the subprocess.

\end{itemize}

The assignment for the numerical values of $g_R({\mu}_0)$,
$g'_R({\mu}_0)$ and ${\alpha}^R_k({\mu}_0)$ parameters appearing in our
formulas has been done as follows:

\begin{itemize}

\item[-] The quantities $g_R$ and $g'_R$, at the chosen scale
${\mu}_0=m_Z$, have been taken from  recent $LEP$ data \cite{datos}.

\item[-] The symmetry breaking pattern that we have considered
corresponds to neglecting all the ${\alpha}^R_k$ parameters at the
${\mu}_0$ scale. We refer to this scenario as the {\it Running Zero
Model}:

\begin{displaymath}
\begin{array}{lccl}
{\alpha}_k ({\mu}_0) & = &  0  &  \;\;  (k=0,1,2,3,4,5)           \\
{\alpha}_k           & = &  0  &  \;\;  (k=6,7,8,9,10,11,12,13)
\end{array}
\end{displaymath}

\end{itemize}

After all these changes in the program, we
apply the different sets of optimal cuts previously obtained,
(Table 3)  and calculate the total number of events to
produce $Z^0Z^0$ or $W^{\pm}Z^0$. The corresponding results
are displayed in Table \ref{tab:run}. We present for each pair of
optimal
cuts, the total number of $Z^0Z^0$ or $W^{\pm}Z^0$ events, in the
{\it Zero Model} and with $g$ and $g'$ constant
($N^{(i)}({\alpha}^0; g,g',c)$), and
in the {\it Running Zero Model} with $g_R(\sqrt{\hat{s}})$ and
$g'_R(\sqrt{\hat{s}})$
($N^{(i)}({\alpha}^0(\sqrt{\hat{s}});
g_R(\sqrt{\hat{s}}), g'_R(\sqrt{\hat{s}}),
c)$). We also give an estimation of the
statistical
significance or the number of sigmas, $r^{(i)}_{\rm run.}$,
corresponding to
include the dependence on the energy of the parameters, with respect to
the tree level results. Here, we calculate $r^{(i)}_{\rm run.}$ function
analogously to eqs.(\ref{eq:rk}) and (\ref{eq:rstruc})

\begin{equation}
r^{(i)}_{\rm run.}(c)  = \frac{| N^{(i)}({\alpha}^0; g,g',c)
- N^{(i)}({\alpha}^0(\sqrt{\hat{s}}); g(\sqrt{\hat{s}}),
g'(\sqrt{\hat{s}}),c) |}
{\sqrt{N^{(i)}({\alpha}^0; g,g',c)}}
\label{eq:rrun}
\end{equation}

As it can be observed in Table \ref{tab:run}, the $r^{(i)}_{\rm run.}$
function, that reflects
the relevance of this effect, is very high when we use the minimal cuts.
Besides, with all  the optimal cuts, the statistical
significance function $r^{(i)}_{\rm run.}$ whose values go from $1$
to $4$ standard deviations is also important. Thus, the effect of the running
 of the
couplings is important and it should be taken into account if one wants
to compute
the total number of $Z^0Z^0$ or $W^{\pm}Z^0$ events.

Now one could ask how the $r_k$ that were  previously obtained (Table
\ref{tab:opti2})
would change if we used the improved tree level approximation with the
running parameters, instead of working only at tree level order.
In sight of the results contained in Table \ref{tab:run} and if we
supposed linear behaviour we would obtain fluctuations in the
statistical significance functions, $r_k$, varying between $2$ to $15\%$
with respect to their values calculated without including the dependence
on the energy of the parameters.

We can summarize these results  saying that the inclusion of
the running coupling constants in our calculations leads to important
variations in the total number of events, given by $r^{(i)}_{\rm run.}$
functions. However, this effect is not so relevant when we want to
determine the $r_k$ functions corresponding to each ${\alpha}_k$,
since the differences obtained with respect to the tree level results
are small. Therefore, the running of the couplings does not change
significantly
our previous discussion about which couplings will be
{\it measurable} at the $LHC$.

\section{Conclusions}

Using $\chi PT$ to describe the $ESB$
sector, we have studied the different processes contributing to $Z^0Z^0$
and $W^{\pm}Z^0$ final states at the $LHC$, considering only gold-plated
events. The main
novelty in our analysis is that we do not use the $ET$ but we have worked
 explicitly with all
the the gauge boson polarization states. Thus we can study the low
energy region, after having imposed a maximal bound on the subprocess
energy of $1.5TeV$.
We have elaborated a $FORTRAN$ code that generates  the subprocess cross
sections at tree level order, producing $Z^0Z^0$ or $W^{\pm}Z^0$ pairs.

With this code we have carried out a systematical study of
the possibilities  for measuring the chiral parameters
${\alpha}_k$, at the $LHC$. This analysis includes an optimization
procedure, that makes possible to  obtain the greatest statistical significance
for measuring  certain values of these parameters. We have also computed
their corresponding statistical errors, $(\Delta\alpha_k)_{\rm stat.}$,
and set the
minimum values of these ${\alpha}_k$ to be unambiguously detected with a
certain statistical significance at the $LHC$. The results of
our studies are displayed in Tables \ref{tab:opti2} and \ref{tab:error}.
{}From our results it is clear that the
 ${\alpha}_0$, $\alpha_1$ and $\alpha_2$ parameters cannot be probed at
the $LHC$. On the contrary, as it can be seen in Tables \ref{tab:opti2}
and \ref{tab:error}, the $r_k$ function obtained for
$\alpha_4$ and $\alpha_5$ can be, for some channels, greater than $3$ sigmas
($2$ sigmas for
$\alpha_3$).
In fact,
 the parameters that could be more easily measured in the $LHC$ are
$\alpha_4 < 0$ in $W^{\pm}Z^0$ and $\alpha_5 > 0$ in $Z^0Z^0$ channel.
As it can be expected their corresponding statistical errors are the
smallest. In all  our calculations, we have fixed a running time corresponding
to
one full $LHC$ year according to the nominal luminosity $L=10^{34}
cm^{-2} sec^{-1}$
($L_{\rm int.}=3 \times 10^5 pb^{-1}$).
The results corresponding to different integrated luminosities can be
obtained just  rescaling the statistical significance by the square root of the
running time.

On the other hand, we have also estimated the size of the imprecisions in our
calculations due to the indetermination in the structure functions by
choosing three different distribution functions. The
results (which can be found in Table \ref{tab:festru}) show that the
errors coming from this effect are smaller than the statistical errors
 obtained for the
$\alpha$ parameters. Thus our present ignorance about the parton distribution
functions does
not restrict the measurement of the chiral parameters.

We have also tried to improve our tree level results considering the dependence
on the
subprocess energy of the weak couplings and the chiral parameters. The size of
these
effects is shown in Table \ref{tab:run}. From our results it is clear that the
total number of
events changes when this effect is taken into account. However, our previous
conclusions
about which parameters can be probed at the $LHC$, do not change at all.

Moreover one could ask about the possibility of increasing the
number of measurable chiral
parameters by relaxing some
of the experimental assumptions of
this work. For instance we could also take into account the gauge boson
hadronic decays,
separate the final polarizations, etc... In that case
we could certainly enhance the statistical significance
 and more chiral parameters could
be probed. However it is not
possible today to have any idea about how well this new information
will be obtained at the
$LHC$ and for this reason we have not considered such possibilities.

  Finally we have also assumed that
the $Z^0Z^0$ or $W^{\pm}Z^0$ pairs produced via electroweak gauge boson
fusion cannot be separated experimentally
 from those coming from another sources such as $gg$ fusion (in
$Z^0Z^0$ case) or quark-antiquark annihilation. Nevertheless  some
forward calorimeters could, presumably, be incorporated to the $LHC$ detectors.
 This fact could allow for a jet tagging at a certain level of
efficiency. This experimental improvement is a realistic way to increase
the statistical significance functions $r_k$ to measure the chiral
Lagrangian parameters at $LHC$.
Work is in progress to treat this possibility and to analyze how much the
sensibility to the $\alpha_k$ parameters could be enhanced.

\section{Acknowledgements}
We thank  M. J. Herrero for her patient help along the more than two years that
this
work took to be finished, as well as D. Denegri, T. Rodrigo, J. Terr\'on for
some interesting
discussions and J. R. Pel\'aez for reading the manuscript. We also thank
support
by the Ministerio de Educaci\'on y Ciencia (Spain) (CICYT AEN90-0034,
AEN93-0776 and AEN93-0673).
A. D. thanks to the CERN Theory Division,
where the final part of this work was done, for its kind hospitality.

\newpage

\section*{Appendix}
\hspace*{12pt}

In this Appendix, we write some of the Feynman rules obtained from the
quantized chiral electroweak Lagrangian ${\cal L}_{\rm NLSM}$
eq.(\ref{eq:MENL}), described in section 2.

The different contributions to ${\cal L}_{\rm NLSM}$ are
given in  eqs.(\ref{eq:lag1}, \ref{eq:YM}, \ref{eq:lag2} and
\ref{eq:quant}), where the Landau gauge has been chosen.

We show the Feynman rules corresponding to the propagators and vertices
used in our calculations:

\hspace{0.5 cm} i) The $\pi$ and $W^{\pm}$ propagators in the Landau
gauge have the following expressions (see Figure \ref{fig:GF2}):

\begin{eqnarray}
-i\Delta({\pi}^{\pm}) & = &
\frac{i}{k^2}   \nonumber   \\
i\Delta_{\mu\nu}(W^{\pm}) & = & \frac{-i}{k^2-m_W^2+i\epsilon}
[g_{\mu\nu}-\frac{k_{\mu}k_{\nu}}{k^2}]
\label{eq:GF2}
\end{eqnarray}

\hspace{0.5 cm} ii) In the following, we write the Feynman rules
corresponding to the vertices with either 3 gauge electroweak bosons
(${\rm I}_{V_1V_2V_3}$) or to 2 gauge bosons and a $GB$ (${\rm
I}_{V_1V_2\pi}$) displayed in Figure \ref{fig:GF3}.

\begin{eqnarray}
{\rm I}_{\pi^+W^-_{\mu}Z^0_{\nu}}          &  =  &
\frac{g^2 v}{2} {\rm sec}{\theta}_W g_{\mu\nu}({\rm sin}^2 {\theta}_W -
2g^2{\beta}_1) + \frac{2 g^2}{v} [(q.r)g_{\mu\nu} - q_{\mu}r_{\nu}]
({\alpha_1}{\rm sin}{\theta}_W
- {\alpha_8}{\rm cos}{\theta}_W +      \nonumber  \\
 & &  {\alpha}_{13}{\rm sec}{\theta}_W) + \frac{2g^2}{v}
{\alpha_3}{\rm sec}{\theta}_W
[(k.r)g_{\mu\nu} - k_{\mu}r_{\nu}] + \frac{2g^2}{v}
[(k.q)g_{\mu\nu}-q_{\mu}k_{\nu}]({\alpha}_2 {\rm sin}{\theta}_W -
\nonumber   \\
& & ({\alpha}_3+{\alpha}_9){\rm cos}{\theta}_W) + \frac{2 g^2}{v}
{\alpha}_{11}
{\rm cos}{\theta}_W [k_{\mu}q_{\nu} - r_{\mu}k_{\nu}] -
\frac{2g^2}{v}{\alpha}_{11} {\rm sin}^2 {\theta}_W {\rm sec} {\theta}_W
[k^2g_{\mu\nu} -      \nonumber    \\
 &  & k_{\mu}q_{\nu}+r_{\mu}r_{\nu}] +
\frac{g^2}{v}{\alpha}_{12}{\rm sec}{\theta}_W
[k^2g_{\mu\nu}-k_{\mu}q_{\nu}+r_{\mu}r_{\nu}+q_{\mu}q_{\nu}] -
\frac{2g^2}{v} {\alpha}_{13}{\rm sec}{\theta}_W q_{\mu}q_{\nu}
\nonumber \\
{\rm I}_{\pi^+W^-_{\mu}A_{\nu}}            &  =  &
-\frac{g^2 v}{2}{\rm sin}^2 {\theta}_W g_{\mu\nu} - \frac{2 g^2}{v}
[(q.r)g_{\mu\nu}-q_{\mu}r_{\nu}]({\alpha_1}{\rm cos}{\theta}_W
+ {\alpha_8}{\rm sin}{\theta}_W) - \frac{2
g^2}{v}[(k.q)g_{\mu\nu}  \nonumber  \\
 &  &  -q_{\mu}k_{\nu}]({\alpha}_2 {\rm cos}{\theta}_W +
({\alpha}_3+{\alpha}_9){\rm sin}{\theta}_W) + \frac{2 g^2}{v}
{\alpha}_{11}
{\rm sin}{\theta}_W [k^2g_{\mu\nu}-k_{\nu}r_{\mu}+r_{\mu}r_{\nu}]
\nonumber   \\
{\rm I}_{W^-_{\mu}W^+_{\nu}Z^0_{\lambda}}  &  =  &
ig{\rm cos}{\theta}_W (1 + {\alpha_3} g^2
{\rm sec}^2{\theta}_W) [(k-r)_{\lambda}
g_{\mu\nu} + (r-q)_{\mu}g_{\nu\lambda} + (q-k)_{\nu} g_{\mu\lambda}] +
ig^2 \{ g{\alpha}_3    \nonumber   \\
  &  &  {\rm sin} {\theta}_W {\rm tg}{\theta}_W + g
({\alpha}_2-{\alpha}_1){\rm sin}{\theta}_W + g({\alpha}_8 -
{\alpha}_9){\rm cos}{\theta}_W - {\alpha}_{13}g {\rm sec}{\theta}_W \}
[q_{\mu}g_{\nu\lambda} - q_{\nu}   \nonumber   \\
 &  &  g_{\mu\lambda}]
+ i g^2 \{ -g{\alpha}_{11} {\rm sin}{\theta}_W {\rm tg}{\theta}_W
+ \frac{{\alpha}_{12}}{2}g
{\rm sec}{\theta}_W \} [k_{\mu}g_{\nu\lambda} - r_{\nu}g_{\mu\lambda}]
\nonumber   \\
{\rm I}_{W^-_{\mu}W^+_{\nu}A_{\lambda}}    &  =  &
ig{\rm sin}{\theta}_W [(k-r)_{\lambda}
g_{\mu\nu} + (r-q)_{\mu}g_{\nu\lambda} + (q-k)_{\nu} g_{\mu\lambda}] +
ig^3 \{ -{\alpha}_3{\rm sin}{\theta}_W +
({\alpha}_1-   \nonumber   \\
& &  {\alpha}_2){\rm cos}{\theta}_W + ({\alpha}_8 -
{\alpha}_9){\rm sin}{\theta}_W  \}
[q_{\mu}g_{\nu\lambda} - q_{\nu}g_{\mu\lambda}] + i g^3
{\alpha}_{11} {\rm sin}{\theta}_W
[k_{\mu}g_{\nu\lambda}  \nonumber  \\
& &  - r_{\nu}g_{\mu\lambda}]
\label{eq:GF3}
\end{eqnarray}

\hspace{0.5 cm} iii) Here, we write the Feynman rules
corresponding to the different vertices with 4 gauge electroweak
bosons
(${\rm I}_{V_1V_2V_3V_4}$), shown in Figure \ref{fig:GF4}.

\begin{eqnarray}
{\rm I}_{W^-_{\mu}W^-_{\nu}W^+_{\lambda}W^+_{\rho}}    &  =  &
i g^2 \{ 1 + g^2({\alpha}_4 - {\alpha}_8) + 2 g^2 ({\alpha}_3 +
{\alpha}_9 + {\alpha}_{13}) [2g_{\mu\nu}g_{\lambda\rho} -
g_{\mu\lambda}g_{\nu\rho} - g_{\mu\rho}g_{\nu\lambda}] + 2ig^4
\nonumber      \\
 &  &  ({\alpha}_4 + {\alpha}_5) [g_{\mu\lambda}g_{\nu\rho} +
g_{\mu\rho}g_{\nu\lambda}]  \nonumber   \\
{\rm I}_{Z^0_{\mu}Z^0_{\nu}W^-_{\lambda}W^+_{\rho}}    &  =  &
i g^2 \{ -{\rm cos}^2{\theta}_W  - 2g^2{\alpha}_3 + g^2{\rm sec}^2{\theta}_W
({\alpha}_5 + {\alpha}_7) \}[2g_{\mu\nu}g_{\lambda\rho} -
g_{\mu\lambda}g_{\nu\rho} - g_{\mu\rho}g_{\nu\lambda}] + ig^4 \nonumber \\
&  &  \{ ({\alpha}_4
+ {\alpha}_5 + {\alpha}_6 + {\alpha}_7) {\rm sec}^2{\theta}_W +
{\alpha}_{12} {\rm tg}^2{\theta}_W -
{\alpha}_{11}{\rm sin}^2{\theta}_W{\rm tg}^2{\theta}_W \}
[g_{\mu\lambda}g_{\nu\rho}+ g_{\mu\rho}g_{\nu\lambda}]  \nonumber  \\
{\rm I}_{A_{\mu}Z^0_{\nu}W^-_{\lambda}W^+_{\rho}}      &  =  &
-i g^2 {\rm sin}{\theta}_W {\rm cos}{\theta}_W [1 + g^2{\rm sec}^2{\theta}_W
{\alpha}_3]  [2g_{\mu\nu}g_{\lambda\rho} -
g_{\mu\lambda}g_{\nu\rho} - g_{\mu\rho}g_{\nu\lambda}] + ig^4
\{ {\rm sin}^2{\theta}_W   \nonumber    \\
 &  &  {\rm tg}{\theta}_W{\alpha}_{11} -
\frac{1}{2} {\rm tg}{\theta}_W
{\alpha}_{12} \} [g_{\mu\lambda}g_{\nu\rho}+ g_{\mu\rho}g_{\nu\lambda}]
\nonumber  \\
{\rm I}_{A_{\mu}A_{\nu}W^-_{\lambda}W^+_{\rho}}        &  =  &
-i g^2  \{ {\rm sin}^2{\theta}_W [2 g_{\mu\nu}g_{\rho\lambda} -
g_{\mu\lambda}g_{\nu\rho} - g_{\mu\rho}g_{\nu\lambda}] + g^2
{\rm sin}^2{\theta}_W
{\alpha}_{11} [ g_{\mu\lambda}g_{\nu\rho} +
g_{\mu\rho}g_{\nu\lambda} ] \}    \nonumber    \\
{\rm I}_{Z^0_{\mu}Z^0_{\nu}Z^0_{\lambda}Z^0_{\rho}}    &  =  &
2i{\rm sec}^4{\theta}_W g^4 [({\alpha}_4 + {\alpha}_5) + 2 ({\alpha}_6 +
{\alpha}_7 + {\alpha}_{10})] [g_{\mu\nu}g_{\lambda\rho} +
g_{\mu\lambda}g_{\nu\rho} + g_{\mu\rho}g_{\nu\lambda}]
\label{eq:GF4}
\end{eqnarray}

\newpage
\section*{Table Captions}
\hspace*{12pt}

\noindent Table \ref{tab:equiv}. In this Table we display the relations
between different sets
of chiral Lagrangian parameters. They correspond to those used by
Longhitano \cite{Long},
Feruglio \cite{Fer} and Dobado et.al. \cite{HeE2}, in
columns $1$, $2$ and $3$, respectively.

\bigskip

\noindent Table \ref{tab:line}.
In this table we show
the sensitivity functions $s^{(ZZ)}_k$ and
$s^{(WZ)}_k$ associated to each ${\alpha}_k$ ($k=0,1,2,3,4,5$)
parameter. They were calculated with the
minimal cuts and under the hypothesis of linear behaviour of
$N^{(i)}({\alpha};c)$ with respect to each ${\alpha}_k$.

\bigskip

\noindent Table \ref{tab:opti1}.
Optimal cuts to calculate the statistical significance
$r^{(i)}_k(c)$ for
typical positive and negative ${\alpha}_k$ values (${\alpha}_k = \pm
0.005$) in the $Z^0Z^0$ or $W^{\pm}Z^0$ final states. The first entry
in column 3 refers to the minimum invariant mass and the second  to
the minimum allowed $p_{TZ}$,
both  them given in $GeV$. The total number of events in
the {\it Zero Model} and with the typical chosen ${\alpha}_k$ obtained
after
having applied the optimal cuts, are represented in columns $4$ and
$5$. They correspond to $1$ year of running for the $LHC$ and a maximal
rapidity of $2.5$ for the final bosons.

\bigskip

\noindent Table \ref{tab:opti2}.
Here, we shown a comparison between the
statistical significance function $r^{(i)}_k(c)$
given by
the typical ${\alpha}_k$ in $Z^0Z^0$ or $W^{\pm}Z^0$ final state, for
the minimal
and the optimal cuts. In both cases, we have fixed the maximal rapidity
cut of $2.5$ for two gauge final bosons, and $1$ year of working time
for the $LHC$.

\bigskip

\noindent Table \ref{tab:error}.
In column 3,  we display the
statistical errors, $(\Delta \alpha_k)_{\rm stat.}$, corresponding
to each ${\alpha}_k$ parameter in the $Z^0Z^0$ or $W^{\pm}Z^0$
channels. They were obtained with the
optimal
cuts (column 1) and under the assumption of linear dependence of the
total number of events on ${\alpha}_k$.

\bigskip

\noindent Table \ref{tab:festru}.
Comparison of the total number of $Z^0Z^0$ or $W^{\pm}Z^0$ events
obtained in the {\it Zero Model} applying the optimal cuts and
choosing
different structure functions: $EHLQ$, $MRSD-$ and $GRVHO$. In the
last three columns, we represent the statistical significance to
estimate
these imprecisions, $r^{(i)}_{\rm struc.}(c)$ eq.(\ref{eq:rstruc}),
between $EHLQ$ and $MRSD-$ (column $6$), $EHLQ$ and
$GRVHO$ ($7$ column) and $MRSD-$ and $GRVHO$ ($8$ column). These
uncertainties are called, respectively, $r^{(i)}_{12}$, $r^{(i)}_{13}$
and $r^{(i)}_{23}$.
In all  these computations we have taken $y_{1 {\rm max}}=y_{2 {\rm
max}}=2.5$ and an integrated luminosity for the $LHC$ of $3 \times
10^5 pb^{-1}$.

\bigskip

\noindent Table \ref{tab:run}.
In this Table we represent the total number of $Z^0Z^0$ or $W^{\pm}Z^0$
pairs obtained at $LHC$ using the optimal cuts and setting
$y_{1 {\rm max}}=y_{2 {\rm
max}}=2.5$ and an integrated luminosity for the $LHC$ of
$3 \times 10^5 pb^{-1}$.
In columns 2 and 3 we display the results in the {\it Zero Model} and
with constant $g$ and $g'$ couplings. Again in columns 4
and 5 we show the
total number of $Z^0Z^0$ or $W^{\pm}Z^0$ events but in the
{\it Running Zero Model} and with $g(\sqrt{\hat{s}})$ and
$g'(\sqrt{\hat{s}})$, eq.(\ref{eq:gsR}).
In the last two
columns we display a measure of the statistical significance that corresponds
to include
the dependence on the energy of the parameters and couplings, given by
the $r^{(ZZ)}_{\rm run.}(c)$ or $r^{(WZ)}_{\rm run.}(c)$  functions,
eq.(\ref{eq:rrun}).

\newpage
\section*{Figure Captions}
\hspace*{12pt}

\noindent Figure \ref{fig:GF2}. Propagators of the $\pi^{\pm}$ and
$W^{\pm}$.

\bigskip

\noindent Figure \ref{fig:GF3}. Vertices with three gauge bosons
or two gauge bosons and one $GB$, obtained from ${\cal L}_{\rm NLSM}$
eq.(\ref{eq:MENL}).

\bigskip

\noindent Figure \ref{fig:GF4}. Different four gauge boson vertices
that appear in the theory described by the lagrangian ${\cal L}_{\rm
NLSM}$  eq.(\ref{eq:MENL}).

\bigskip

\noindent Figure \ref{fig:gg}. Different diagrams to one loop order
contributing to $gg \rightarrow Z^0Z^0$ cross section, in the $SM$
without a Higgs boson.

\bigskip

\noindent Figure \ref{fig:qq}.  We display the diagrams at tree level order
that
contribute to  the cross section $q\bar{q} \rightarrow Z^0Z^0$ in the
framework of
chiral perturbation theory (with ${\cal L}_{\rm NLSM}$).
In this case we obtain the same result as in the $SM$ without a Higgs boson.

\bigskip

\noindent Figure \ref{fig:ZZ}.  We show  the only diagram contributing
to the
different $Z^0Z^0 \rightarrow Z^0Z^0$ helicity amplitudes obtained with
${\cal L}_{\rm NLSM}$
eq.(\ref{eq:MENL}), at tree level order.

\bigskip

\noindent Figure \ref{fig:WW}. The scattering amplitudes for the
process $W^+W^- \rightarrow Z^0Z^0$, calculated with ${\cal L}_{\rm
NLSM}$ eq.(\ref{eq:MENL}) at tree level order in the Landau gauge,
receive contribution from all these channels.

\bigskip

\noindent Figure \ref{fig:qq'}. The cross sections $q\bar{q'} \rightarrow
W^{\pm}Z^0$ are calculated adding the contribution of the diagrams
displayed in this Figure. They are obtained with ${\cal L_{\rm NLSM}}$
eq.(\ref{eq:MENL}) at tree level order and in the Landau gauge. The
initial quarks are supposed to be massless.

\bigskip

\noindent Figure \ref{fig:WZ}. Different diagrams contributing to the
$W^{\pm}Z^0 \rightarrow W^{\pm}Z^0$ scattering amplitudes. They are calculated
with ${\cal L}_{\rm NLSM}$ eq.(\ref{eq:MENL}) at tree level order and
in the Landau gauge.

\bigskip

\noindent Figure \ref{fig:Wf}. The scattering amplitudes
$W^{\pm}\gamma \rightarrow W^{\pm}Z^0$
are obtained with the Lagrangian given in eq.(\ref{eq:MENL})  adding
the contribution of the diagrams
displayed in this Figure. The calculation is made at tree level order
and fixing the Landau gauge.

\bigskip

\noindent Figure \ref{fig:sena0}. This graphic represents the
sensitivity of the
$LHC$ to the ${\alpha}_0$ parameter in $Z^0Z^0$ and $W^{\pm}Z^0$
channels. We display the total number of $Z^0Z^0$ (solid line) and $W^{\pm}Z^0$
(dashed
line) events obtained in the $LHC$, for different values of the
${\alpha}_0$ parameter. The kinematical cuts are the
minimal ones.
The other ${\alpha}_k$ parameters, apart from
${\alpha}_0$, have been set to zero.

\bigskip

\noindent Figure \ref{fig:sena1}. This Figure represents the sensitivity
function to
${\alpha}_1$ parameter for the $LHC$, using the same kinematical cuts as those
in
Figure \ref{fig:sena0}. We display the total  $N^{(ZZ)}$ (solid line)
and
$N^{(WZ)}$ (dashed line) events obtained versus ${\alpha}_1$ values.

\bigskip

\noindent Figure \ref{fig:sena2}. This Figure represents the sensitivity
function to
${\alpha}_2$ parameter for the $LHC$. As in Figures \ref{fig:sena0} and
\ref{fig:sena1} the total number of $Z^0Z^0$ and
$W^{\pm}Z^0$
events versus ${\alpha}_2$ values.

\bigskip

\noindent Figure \ref{fig:sena3}. This Figure represents the sensitivity
function to
${\alpha}_3$ parameter for the $LHC$, using the same kinematical cuts as in
previous Figures. We show the total  $N^{(ZZ)}$ (solid line) and
$N^{(WZ)}$ (dashed line) events versus ${\alpha}_3$.

\bigskip

\noindent Figure \ref{fig:sena4}. This graphic represents the
sensitivity of the $LHC$ to the ${\alpha}_4$
parameter in the $Z^0Z^0$ and $W^{\pm}Z^0$ channels. We display,  like
in Figures
\ref{fig:sena0}, \ref{fig:sena1}, \ref{fig:sena2} and \ref{fig:sena3},
the total number of $Z^0Z^0$ (solid line) and $W^{\pm}Z^0$
(dashed line) events obtained at $LHC$, for different values of the
${\alpha}_4$ parameter, using the same kinematical cuts as in
previous Figures.

\bigskip

\noindent Figure \ref{fig:sena5}. Sensitivity function to
${\alpha}_5$ parameter for the $LHC$. We have used the same kinematical cuts as
in
previous Figures. We display the total  $N^{(ZZ)}$ (solid line) and
$N^{(WZ)}$ (dashed line) events obtained versus ${\alpha}_5$ values, as we did
in figures
\ref{fig:sena0} to
\ref{fig:sena4}.

\bigskip

\noindent Figure \ref{fig:mvcut}. In this Figure we study the variation
of the total number of $Z^0Z^0$ (solid line)
and $W^{\pm}Z^0$ (dashed line) events with the cut on the
minimal invariant mass, $\sqrt{\hat{s}_{\rm min}}$. We have set all the
${\alpha}_k$ parameters
to their values in the {\it Zero Model} and the cuts
${p_{TZ}}_{\rm min}$, ${y_1}_{\rm max}$ and ${y_2}_{\rm max}$ to their
minimal values.

\bigskip

\noindent Figure \ref{fig:ptcut}. As in Figure
\ref{fig:mvcut}, we display the behaviour of the total number
of $Z^0Z^0$ (solid line) and $W^{\pm}Z^0$ (dashed line) events
in the {\it Zero Model} versus the cut in the minimal transversal
momentum of the $Z^0$ boson. We have taken the other kinematical bounds
as their minimal values.

\bigskip

\noindent Figure \ref{fig:yicut}. Here we reflect
the dependence of the total
number of  $Z^0Z^0$ and $W^{\pm}Z^0$ events on the maximal rapidity
bounds of
final gauge bosons, $y_{1 {\rm max}} = y_{2 {\rm max}}$.
We work in the {\it Zero Model}  and fix the minimal cuts in
invariant mass and transversal moment. The solid line
corresponds to the $Z^0Z^0$ final state and the dashed one to the produced
$W^{\pm}Z^0$ pairs.

\bigskip

\noindent Figure \ref{fig:siga0}. Statistical significance corresponding
to the ${\alpha}_0$ parameter. We display
the $r_0$ function eq.(\ref{eq:rk}) for different
values of ${\alpha}_0$ with respect to the {\it Zero Model}.
The solid line represents the $Z^0Z^0$ final state and the dashed line the
$W^{\pm}Z^0$. We use the same kinematical cuts as in
Figures
\ref{fig:sena0} to \ref{fig:sena5}.

\bigskip

\noindent Figure \ref{fig:siga1}. Statistical significance corresponding
to the
${\alpha}_1$ parameter. We show as in  Figure \ref{fig:siga0}, the
$r^{(ZZ)}_1$ (solid) eq.(\ref{eq:rk})  and
$r^{(WZ)}_1$ (dashed) functions with respect to the {\it Zero Model} versus
these
${\alpha}_1$ values. We use the same kinematical cuts as in Figures
\ref{fig:sena0} to \ref{fig:sena5}.

\bigskip

\noindent Figure \ref{fig:siga2}. This Figure represents the statistical
significance
for some values of ${\alpha}_2$ with respect to the {\it Zero Model}.
We display as in
Figures \ref{fig:siga0} and \ref{fig:siga1},
and for the same kinematical cuts, the $r^{(ZZ)}_2$  and $r^{(WZ)}_2$
eq.(\ref{eq:rk}) functions.

\bigskip

\noindent Figure \ref{fig:siga3}. This Figure reflects the statistical
significance of ${\alpha}_3$. We show  the $r_k$  eq.(\ref{eq:rk})
functions for $Z^0Z^0$ (solid) and $W^{\pm}Z^0$
(dashed) with respect to the {\it Zero Model} versus ${\alpha}_3$ values,
using the same kinematical cuts as in previous Figures.

\bigskip

\noindent Figure \ref{fig:siga4}. Statistical significance corresponding
to the ${\alpha}_4$ parameter. As in Figures
\ref{fig:siga0} to \ref{fig:siga3}  we show
the $r^{(ZZ)}_4$  (solid) (eq. \ref{eq:rk})  and
$r^{(WZ)}_4$ (dashed) with respect to the {\it Zero Model} versus
the ${\alpha}_4$ values. We have used the same kinematical cuts as in
previous figures.

\bigskip

\noindent Figure \ref{fig:siga5}. The same as in Figures \ref{fig:siga0}
to \ref{fig:siga4} for the  ${\alpha}_5$ parameter.

\clearpage


\begin{table}[t]
\caption{}
\vspace{0.7cm}
\centering
\begin{tabular}{|c|c|c|} \hline\hline
${\alpha}_0$       & $a_0/g^2$     & $\delta\beta/g^2$
\\ \hline
${\alpha}_1$      & $g'/g a_1$    & $L_{10}$tg${\theta}_W$   \\
\hline
${\alpha}_2$      & $g'/g a_2$    &
$-L_{9R}/2$tg${\theta}_W$   \\   \hline
${\alpha}_3$      & $-a_3$               &
$-L_{9L}/2$                    \\   \hline
${\alpha}_4$      & $a_4$                & $L_2$   \\ \hline
${\alpha}_5$      & $a_5$                & $L_1$   \\ \hline
${\alpha}_6$      & $a_6$                & $\;-\;$   \\ \hline
${\alpha}_7$      & $a_7$                & $\;-\;$   \\ \hline
${\alpha}_8$      & $-a_8$               & $\;-\;$   \\ \hline
${\alpha}_9$      & $-a_9$               & $\;-\;$   \\ \hline
${\alpha}_{10}$   & $2a_{10}$            & $\;-\;$   \\ \hline
${\alpha}_{11}$   & $a_{11}$             & $\;-\;$   \\ \hline
${\alpha}_{12}$   & $2a_{12}$            & $\;-\;$   \\ \hline
${\alpha}_{13}$   & $a_{13}$             & $\;-\;$   \\ \hline
\end{tabular}
\label{tab:equiv}
\end{table}

\begin{table}[h]
\caption{}
\vspace{0.7cm}
\centering
\begin{tabular}{|c|r|r|} \hline\hline
${\alpha}_k$      & $s^{(ZZ)}_k(c^{\rm min})$
& $s^{(WZ)}_k(c^{\rm min})$    \\ \hline
${\alpha}_0$      & $-899.54$      & $2779.89$       \\  \hline
${\alpha}_1$      & $1241.27$      & $-5408.20$      \\  \hline
${\alpha}_2$      & $-1481.65$     & $6150.98$       \\  \hline
${\alpha}_3$      & $6751.24$      & $-48366.79$     \\  \hline
${\alpha}_4$      & $2445.43$      & $-18981.46$     \\  \hline
${\alpha}_5$      & $5924.90$      & $-20142.33$     \\  \hline
\end{tabular}
\label{tab:line}
\end{table}

\clearpage

\begin{table}[t]
\caption{}
\vspace{0.7cm}
\centering
\begin{tabular}{|l|c|c|c|c|} \hline\hline
${\alpha}_k$ value & Channel & ${c}^{\rm op} =
(\sqrt{\hat{s}^{\rm op}_{\rm min}}, p^{\rm op}_{T {\rm min}})$   &
$N^{(i)} ({\alpha}^0; {c}^{\rm op})$  &
$N^{(i)} ({\alpha}; {c}^{\rm op})$    \\  \hline
${\alpha}_3 = -0.005$  & $Z^0Z^0$ & $(200,10)$ & 16152.55  &  16119.28
\\    \hline
${\alpha}_3 = -0.005$  & $W^{\pm}Z^0$ & $(200,300)$ & 522.59    & 581.41
\\    \hline
${\alpha}_3 =  0.005$  & $Z^0Z^0$ & $(200,10)$ & 16152.55   & 16186.47
\\    \hline
${\alpha}_3 =  0.005$  & $W^{\pm}Z^0$ & $(200,200)$ & 2268.62  &
2160.74   \\  \hline
${\alpha}_4 = -0.005$  & $W^{\pm}Z^0$ & $(1150,500)$ & $32.29$ & $66.48$
\\    \hline
${\alpha}_4 =  0.005$  & $Z^0Z^0$ & $(1150,400)$ & $13.50$ & $20.01$
\\  \hline
${\alpha}_5 = -0.005$  & $W^{\pm}Z^0$ & $(1150,400)$ & $61.56$ & $85.78$
\\ \hline
${\alpha}_5 =  0.005$  & $Z^0Z^0$ & $(1150,400)$ & $13.50$ & $28.70$
\\ \hline
\end{tabular}
\label{tab:opti1}
\end{table}

\begin{table}[h]
\caption{}
\vspace{0.7cm}
\centering
\begin{tabular}{|l|c|c|c|} \hline\hline
${\alpha}_k$ value & Channel & $r_k({\alpha}; {c}^{\rm min})$  &
$r_k({\alpha}; {c}^{\rm op})$      \\  \hline
${\alpha}_3 = -0.005$  & $Z^0Z^0$ & $0.26$  & $0.26$       \\  \hline
${\alpha}_3 = -0.005$  & $W^{\pm}Z^0$ & $1.01$  & $2.57$   \\  \hline
${\alpha}_3 =  0.005$  & $Z^0Z^0$ & $0.27$  & $0.27$        \\  \hline
${\alpha}_3 =  0.005$  & $W^{\pm}Z^0$ & $0.51$  & $2.27$    \\  \hline
${\alpha}_4 = -0.005$  & $W^{\pm}Z^0$ & $0.45$  & $6.02$    \\  \hline
${\alpha}_4 =  0.005$  & $Z^0Z^0$ & $0.18$  & $1.77$    \\  \hline
${\alpha}_5 = -0.005$  & $W^{\pm}Z^0$ & $0.39$  & $3.09$    \\  \hline
${\alpha}_5 =  0.005$  & $Z^0Z^0$ & $0.37$  & $4.14$    \\  \hline
\end{tabular}
\label{tab:opti2}
\end{table}

\begin{table}[h]
\caption{}
\vspace{0.7cm}
\centering
\begin{tabular}{|c|c|c|} \hline\hline
${c}^{\rm op} =
(\sqrt{\hat{s}^{\rm op}_{\rm min}}, p^{\rm op}_{T {\rm min}})$
& Channel &  $(\Delta {\alpha}_k)_{\rm stat.}$   \\ \hline
$(200,300)$ & $W^{\pm}Z^0$ &  $(\Delta {\alpha}_3)_{\rm stat.} = 1.95
\times   10^{-3}$      \\   \hline
$(200,200)$ & $W^{\pm}Z^0$ &  $(\Delta {\alpha}_3)_{\rm stat.} = 2.20
\times   10^{-3}$      \\   \hline
$(1150,500)$ & $W^{\pm}Z^0$ &  $(\Delta {\alpha}_4)_{\rm stat.} = 8.31
\times   10^{-4}$      \\   \hline
$(1150,400)$ & $Z^0Z^0$ &  $(\Delta {\alpha}_4)_{\rm stat.} = 2.82
\times    10^{-3}$      \\   \hline
$(1150,400)$ & $W^{\pm}Z^0$ &  $(\Delta {\alpha}_5)_{\rm stat.} = 1.62
\times      10^{-3}$      \\   \hline
$(1150,400)$ & $Z^0Z^0$ &  $(\Delta {\alpha}_5)_{\rm stat.} = 1.21
\times    10^{-3}$      \\   \hline
\end{tabular}
\label{tab:error}
\end{table}

\clearpage

\begin{table}[t]
\caption{}
\vspace{0.7cm}
\centering
\begin{tabular}{|c|c|c|c|c|c|c|c|} \hline\hline
Channel & Optimal cuts & $N^{(i)}({\alpha}^0; c^{\rm
op})$ & $N^{(i)}({\alpha}^0; c^{\rm op})$
& $N^{(i)}({\alpha}^0; c^{\rm op})$ & $r^{(i)}_{12}$
& $r^{(i)}_{13}$  & $r^{(i)}_{23}$         \\
  & $({\sqrt{\hat{s}_{\rm min}}}^{\rm op}$, $p^{\rm op}_{T \rm min})$
&
$EHLQ$({\it set} $II$) &  $MRS -D'$ & $GRV HO$ &  & &    \\
 &  &  &  &  &  &  &  \\  \hline
$ZZ$ & ($200,10$) & 16113.31 & 17242.93 & 16617.27 & 8.90  & 3.97 &
4.76     \\   \hline
$WZ$ & ($200,200$) & 2298.57  & 2526.75  & 2452.56  & 4.76 & 3.21 & 1.48
\\
\hline
$WZ$ & ($200,300$) & 535.80 & 594.34  & 578.75 & 2.53 & 1.86 & 0.64  \\
\hline
$WZ$ & ($1150,500$) & 36.02 & 40.79 & 39.97 & 0.79 & 0.66 & 0.13 \\
\hline
$ZZ$ & ($1150,400$) & 14.92 & 15.88 & 15.39 & 0.25 & 0.12 & 0.12 \\
\hline
$WZ$ & ($1150,400$) & 68.02 & 76.72 & 74.78 & 1.05 & 0.82 & 0.22 \\
\hline
\end{tabular}
\label{tab:festru}
\end{table}

\begin{table}[h]
\caption{}
\vspace{0.7cm}
\centering
\begin{tabular}{|r|r|r|c|c|c|c|}   \hline
Optimal cuts &  \multicolumn{2}{c|}{No running}  &
\multicolumn{2}{c|}{Running}  &  \multicolumn{2}{c|}{$r^{(i)}_{\rm
run.}$}     \\   \cline{2-7}

$(\sqrt{\hat{s}_{\rm min}})$  &  $N^{(ZZ)}(\alpha^0;c^{\rm op})$     &
$N^{(WZ)}(\alpha^0;c^{\rm op})$  &  $N^{(ZZ)}(\alpha^0;c^{\rm op})$  &
$N^{(WZ)}(\alpha^0;c^{\rm op})$  & $r^{(ZZ)}_{\rm run.}$  &
$r^{(WZ)}_{\rm run.}$     \\  \hline
$(200,10)$ & 16149.13 & 91141.49 & 15355.82 & 86149.13 & 6.24 & 16.54
\\      \hline
$(200,200)$  & $-\;\;\;\;$   & 2281.63  & $\;\;-\;\;$  & 2093.83  &
$\;\;-\;\;$  &  3.93  \\    \hline
$(200,300)$  & $-\;\;\;\;$   & 532.33   & $\;\;-\;\;$  & 505.18   &
$\;\;-\;\;$  & 1.18   \\    \hline
$(1150,500)$ & $-\;\;\;\;$   & 36.01    & $\;\;-\;\;$  & 45.75    &
$\;\;-\;\;$  & 1.62   \\    \hline
$(1150,400)$ & 14.91 & 68.01 & 11.28 & 78.07 & 0.94 & 1.22  \\  \hline
\end{tabular}
\label{tab:run}
\end{table}

\clearpage

\begin{figure}

\caption{}
\label{fig:GF2}
\end{figure}

\clearpage

\begin{figure}

\caption{}
\label{fig:GF3}
\end{figure}

\clearpage

\begin{figure}

\caption{}
\label{fig:GF4}
\end{figure}

\clearpage

\begin{figure}
\vspace{2 cm}

\caption{}
\label{fig:gg}
\end{figure}

\begin{figure}
\vspace{2 cm}

\caption{}
\label{fig:qq}
\end{figure}

\clearpage

\begin{figure}
\vspace{2 cm}
\caption{}
\label{fig:ZZ}
\end{figure}

\clearpage

\begin{figure}

\caption{}
\label{fig:WW}
\end{figure}

\clearpage

\begin{figure}
\vspace{3 cm}

\caption{}
\label{fig:qq'}
\end{figure}

\begin{figure}

\caption{}
\label{fig:WZ}
\end{figure}

\clearpage

\begin{figure}

\caption{}
\label{fig:Wf}
\end{figure}

\clearpage

\begin{figure}

\caption{}
\label{fig:sena0}
\end{figure}

\begin{figure}

\caption{}
\label{fig:sena1}
\end{figure}

\begin{figure}

\caption{}
\label{fig:sena2}
\end{figure}

\begin{figure}

\caption{}
\label{fig:sena3}
\end{figure}

\clearpage

\begin{figure}

\caption{}
\label{fig:sena4}
\end{figure}

\begin{figure}

\caption{}
\label{fig:sena5}
\end{figure}

\clearpage

\begin{figure}

\caption{}
\label{fig:mvcut}
\end{figure}

\begin{figure}

\caption{}
\label{fig:ptcut}
\end{figure}

\clearpage

\begin{figure}

\caption{}
\label{fig:yicut}
\end{figure}

\clearpage

\begin{figure}

\caption{}
\label{fig:siga0}
\end{figure}

\begin{figure}

\caption{}
\label{fig:siga1}
\end{figure}

\clearpage

\begin{figure}
\caption{}
\label{fig:siga2}
\end{figure}

\clearpage

\begin{figure}
\caption{}
\label{fig:siga3}
\end{figure}

\begin{figure}
\caption{}
\label{fig:siga4}
\end{figure}

\clearpage

\begin{figure}

\caption{}
\label{fig:siga5}
\end{figure}

\clearpage

\end{document}